\documentclass[twocolumn]{aastex631}

\usepackage{amsmath}
\usepackage{dirtytalk}
\usepackage{enumitem}
\received{August 29, 2023}
\accepted{December 3, 2023}

\begin{document}

\title{Jet reorientation in central galaxies of clusters and groups: insights from VLBA and Chandra data}

\correspondingauthor{Francesco Ubertosi}
\email{francesco.ubertosi2@unibo.it}

\author[0000-0001-5338-4472]{Francesco Ubertosi}\affiliation{Dipartimento di Fisica e Astronomia, Università di Bologna, via Gobetti 93/2, I-40129 Bologna, Italy}
\affiliation{Istituto Nazionale di Astrofisica - Osservatorio di Astrofisica e Scienza dello Spazio (OAS), via Gobetti 101, I-40129 Bologna, Italy}

\author[0000-0002-4962-0740]{Gerrit Schellenberger}
\affiliation{Center for Astrophysics $|$ Harvard \& Smithsonian, 60 Garden Street, Cambridge, MA 02138, USA}

\author[0000-0002-5671-6900]{Ewan O'Sullivan}
\affiliation{Center for Astrophysics $|$ Harvard \& Smithsonian, 60 Garden Street, Cambridge, MA 02138, USA}

\author[0009-0007-0318-2814]{Jan Vrtilek}
\affiliation{Center for Astrophysics $|$ Harvard \& Smithsonian, 60 Garden Street, Cambridge, MA 02138, USA}

\author[0000-0002-1634-9886]{Simona Giacintucci}
\affiliation{Naval Research Laboratory, 4555 Overlook Avenue SW, Code 7213, Washington, DC 20375, USA}

\author{Laurence P. David}
\affiliation{Center for Astrophysics $|$ Harvard \& Smithsonian, 60 Garden Street, Cambridge, MA 02138, USA}

\author[0000-0002-9478-1682]{William Forman}
\affil{Center for Astrophysics $|$ Harvard \& Smithsonian, 60 Garden Street, Cambridge, MA 02138, USA}

\author[0000-0002-0843-3009]{Myriam Gitti}
\affil{Dipartimento di Fisica e Astronomia, Università di Bologna, via Gobetti 93/2, I-40129 Bologna, Italy}
\affil{Istituto Nazionale di Astrofisica - Istituto di Radioastronomia (IRA), via Gobetti 101, I-40129 Bologna, Italy}

\author[0000-0002-8476-6307]{Tiziana Venturi}
\affil{Istituto Nazionale di Astrofisica - Istituto di Radioastronomia (IRA), via Gobetti 101, I-40129 Bologna, Italy}

\author{Christine Jones}
\affil{Center for Astrophysics $|$ Harvard \& Smithsonian, 60 Garden Street, Cambridge, MA 02138, USA}

\author[0000-0001-9807-8479]{Fabrizio Brighenti}
\affil{Dipartimento di Fisica e Astronomia, Università di Bologna, via Gobetti 93/2, I-40129 Bologna, Italy}
\affil{University of California Observatories/Lick Observatory, Department of Astronomy and Astrophysics, Santa Cruz, CA 95064, USA}

%% Note that the \and command from previous versions of AASTeX is now
%% depreciated in this version as it is no longer necessary. AASTeX 
%% automatically takes care of all commas and "and"s between authors names.

%% AASTeX 6.31 has the new \collaboration and \nocollaboration commands to
%% provide the collaboration status of a group of authors. These commands 
%% can be used either before or after the list of corresponding authors. The
%% argument for \collaboration is the collaboration identifier. Authors are
%% encouraged to surround collaboration identifiers with ()s. The 
%% \nocollaboration command takes no argument and exists to indicate that
%% the nearby authors are not part of surrounding collaborations.

%% Mark off the abstract in the ``abstract'' environment. 
\begin{abstract}
Recent observations of galaxy clusters and groups with misalignments between their central AGN jets and X-ray cavities, or with multiple misaligned cavities, have raised concerns about the jet -- bubble connection in cooling cores, and the processes responsible for jet realignment. To investigate the frequency and causes of such misalignments, we construct a sample of 16 cool core galaxy clusters and groups. 
Using VLBA radio data we measure the parsec-scale position angle of the jets, and compare it with the position angle of the X-ray cavities detected in $Chandra$ data. Using the overall sample and selected subsets, we consistently find that 
there is a 30\% -- 38\% chance to find a misalignment larger than $\Delta\Psi = 45^{\circ}$ when observing a cluster/group with a detected jet and at least one cavity.
We determine that projection may account for an apparently large $\Delta\Psi$ only in a fraction of objects ($\sim$35\%), and given that gas dynamical disturbances (as sloshing) are found in both aligned and misaligned systems, we exclude environmental perturbation as the main driver of cavity -- jet misalignment. Moreover, we find that large misalignments (up to $\sim90^{\circ}$) are favored over smaller ones ($45^{\circ}\leq\Delta\Psi\leq70^{\circ}$), and that the change in jet direction can occur on timescales between one and a few tens of Myr. We conclude that misalignments are more likely related to actual reorientation of the jet axis, and we discuss several engine-based mechanisms that may cause these dramatic changes.
\end{abstract}

%% Keywords should appear after the \end{abstract} command. 
%% The AAS Journals now uses Unified Astronomy Thesaurus concepts:
%% https://astrothesaurus.org
%% You will be asked to selected these concepts during the submission process
%% but this old "keyword" functionality is maintained in case authors want
%% to include these concepts in their preprints.

%% From the front matter, we move on to the body of the paper.
%% Sections are demarcated by \section and \subsection, respectively.
%% Observe the use of the LaTeX \label
%% command after the \subsection to give a symbolic KEY to the
%% subsection for cross-referencing in a \ref command.
%% You can use LaTeX's \ref and \label commands to keep track of
%% cross-references to sections, equations, tables, and figures.
%% That way, if you change the order of any elements, LaTeX will
%% automatically renumber them.
%%
%% We recommend that authors also use the natbib \citetp
%% and \citett commands to identify citations.  The citations are
%% tied to the reference list via symbolic KEYs. The KEY corresponds
%% to the KEY in the \bibitem in the reference list below. 

\section{Introduction} \label{sec:intro}
In the context of groups and clusters of galaxies in the local Universe, it is now clear that the co-evolution of supermassive black holes (SMBH) and their environment is regulated by a feedback mechanism. 
Gas, cooling out of the hot intracluster/intragroup medium (ICM/IGrM), fuels the central active galactic nucleus (AGN), that in turn drives jets and outflows propagating to tens or hundreds of kpc. These outbursts can pierce and mechanically heat the surrounding gas, thereby delaying further cooling (e.g., for reviews \citealt{2003ARA&A..41..191M,2007ARA&A..45..117M,2012ARA&A..50..455F,2012AdAst2012E...6G,2012NJPh...14e5023M,2019SSRv..215....5W,2021Univ....7..142E,2022PhR...973....1D}). The clearest footprints of jet activity on the ICM/IGrM are X-ray-void bubbles (or cavities) inflated in the gas by the expanding jet plasma, which can remain visible in X-ray observations for tens of Myr after the outburst began (e.g., \citealt{2000A&A...356..788C,2007ApJ...659.1153W,2010ApJ...714..758G,2011MNRAS.416.2916O,2015ApJ...805..112R,2020ApJ...891....1G,2021ApJ...923L..25U,2023ApJ...948..101S}). 
\\ The observation and investigation of X-ray cavities in clusters and groups highlights several characteristics of SMBH activity. First, the presence of multiple bubbles in the cluster or group reflects the {\it transitory} and {\it repetitive} nature of radio activity in central galaxies. Many systems have several pairs of cavities at increasing distance from the center (from a few kpc to $\sim10^{2}$ kpc), with each pair related to a jet episode that lasted around $10^{7}$ yr (e.g., \citealt{2004ApJ...607..800B,2005MNRAS.363..891F,2005MNRAS.364.1343D,2007ApJ...659.1153W,2013ApJ...768...11B,2015ApJ...805...35H,2015ApJ...805..112R,2021A&A...650A.170B}). The multiple bubbles are evidence for successive jet -- ICM/IGrM interactions, providing valuable information on the duty cycle and mechanical energy of AGN in dense environments. 
\\Second, the AGN activity is {\it directional}. While the mechanical energy can be distributed in every direction (through e.g. cocoon shock fronts, see \citealt{2006ApJ...643..120B,2022MNRAS.516.3750H}, or internal waves generated by the buoyant cavities, see \citealt{2019AAS...23343807Z}), the bipolar jets that deposited such energy were driven in specific and opposite directions from the SMBH. In this context, in the absence of other effects, the position of bubbles in the cluster or group atmosphere maps the direction of the corresponding radio jets. In several cases, older and younger pairs of X-ray cavities are aligned along a common axis, suggesting that the jet has been pointing in a similar direction over time (e.g., \citealt{2007ApJ...659.1153W,2015ApJ...805..112R}). However, observations of objects with jets pointing to a very different direction than the cavity location, or with multiple X-ray cavities spread over the whole azimuth, support the idea that the jet axis can also change its orientation over time (e.g., \citealt{2010ApJ...713L..74F,2012A&A...545L...3C,2013ApJ...768...11B,2021ApJ...906...16S,2021ApJ...923L..25U}).
Since the connection between the AGN jets and the X-ray cavities is vital to support the idea of an AGN feedback mechanism, several scenarios have been investigated to understand the cause behind these dramatic misalignments.
\\Many possibilities are related to the physics of the SMBH and the conditions of the innermost parsecs of the central galaxy. Intervening obstacles or strong magnetic fields could bend or deflect the jets and change their direction over time (\citealt{2013Sci...339...49M,2022AAS...24010124M}). A periodic precession of the jet axis, possibly caused by the jet-accretion disk interaction, may also cause successive outbursts to appear rotated (\citealt{2018MNRAS.474L..81L,2022ApJ...936L...5L,2022MNRAS.516.3750H}). Alternatively, binary SMBH could explain both a progressive change in the jets axis, due to the interaction between the active SMBH and the orbiting one, and the existence of multiple jets and cavities in different directions, if both SMBHs are active (e.g., \citealt{2006A&A...453..433H,2018JApA...39....8R,2021ApJ...923L..25U}). Verifying these possibilities is non-trivial, as the involved spatial scales can be beyond the reach of radio observations. 
\\ Another possibility is that projection effects can cause a large apparent offset, while the actual misalignment between jets and cavity direction is small (see \citealt{2016ApJ...827...66S,2021ApJ...906...16S}). This explanation may alleviate the difference in relative position of jets and cavities observed in a fraction of systems. 
\\Ultimately, environmental effects might play a major role. Dynamical disturbances can trigger sloshing motions of the hot gas, that oscillates in the potential well of the host cluster or group. Investigations of these motions and X-ray cavities revealed that sloshing may move X-ray cavities around the cluster atmosphere (e.g., \citealt{2010ApJ...713L..74F,2010ApJ...722..825R,2013ApJ...773..114P,2020MNRAS.496.1471K,2021ApJ...914...73Z}), potentially perturbing the bubble motion away from the original axis of the radio jets. 
\\As feedback models rely on the central AGN to balance cooling, the observations of jets, in cool core clusters and groups, misaligned with the X-ray cavities has raised concerns on the connection between the direct evidence (the jets) and the smoking gun (the X-ray cavities) of feedback. Such concerns have typically been addressed by focusing on single peculiar objects, but it is essential to analyze a larger sample of clusters and groups to provide a broader picture. 
\\In this work we present the investigation of jets and X-ray cavity alignment using a sample of galaxy clusters and groups, to assess the relative importance of projection and environmental effects, to evaluate how rare these misalignments are, and to gain new insights on the mechanism through which AGN jets change direction over time. Combining $Chandra$ and Very Long Baseline Array (VLBA) data, we observationally address the alignment of X-ray cavities (on kpc scales) and radio jets (on pc scales). 
\\This paper is organized as follows. Section \ref{sec:select} presents the criteria for the sample selection; Section \ref{sec:data} and \ref{sec:measureangle} describe the data and the methods use to measure position angles of jets and X-ray cavities, and Section \ref{sec:results} shows the results. We discuss our findings in Section \ref{sec:disc}, while Section \ref{sec:concl} summarizes our work. Throughout this work we assume a $\Lambda$CDM cosmology with H$_{0}$=70~km~s$^{-1}$~Mpc$^{-1}$, $\Omega_{\text{m}}=0.3$, and $\Omega_{\Lambda}=0.7$. Uncertainties are reported at the 1$\sigma$ confidence level. We define the radio spectrum as $F_{\nu}\propto\nu^{\alpha}$, where $F_{\nu}$ is the flux density at frequency~$\nu$.
\section{Sample selection} \label{sec:select}
We select groups and clusters from the parent catalog presented in \citet{2014PhDT.......338H,2015MNRAS.453.1223H,2015MNRAS.453.1201H}, where 59 central cluster and group galaxies have been classified by their radio morphology on parsec scales. The fairly uniform VLBA observations at 5~GHz are suited to determine the presence and direction of jets in the central dominant galaxies. Since we are driven by the aim of a {\it Chandra} -- VLBA comparison, we include only the sources that have more than 10ks of observations in the {\it Chandra} archive (45 out of 59 objects). By excluding the objects that do not show evidence for extended radio jets in VLBA data and of X-ray cavities in the $Chandra$ data (see the analysis performed in Sect. \ref{sec:data} and \ref{sec:measureangle}), we obtain a final sample of 16 systems (see Tab. \ref{tab:sample}). We note that all these systems host ionized gas nebulae at their centers (as probed by optical line emission, such as H$\alpha$; see \citealt{2014PhDT.......338H} and references therein), strongly indicating the presence of a cooling core in the host cluster or group.
\\The 59 radio-bright BCGs in the sample of \citet{2015MNRAS.453.1223H,2015MNRAS.453.1201H} do not form a complete sample, but were treated as representative of their parent, complete catalog of over 700 clusters. Our further selection (as detailed above and in Sect. \ref{sec:data}) is mainly based on the presence of at least one jet in the radio images. Compact radio sources may either be pointing directly towards us (meaning that any jet would be projected on the core), or may be in a fading phase (they were driving jets in the past that are no longer visible). Thus, the sources in our sub-sample constitute the BCGs with the most active AGN at their centers, and with a small jet inclination angle to the plane of the sky. Far from being complete, the clusters and groups in our sample (Tab. \ref{tab:sample}) are representative of systems where AGN feedback activity is ongoing and vigorous. This matches our aim to understand how frequently and why the pc-scale and kpc-scale evidence of feedback are misaligned.

%%%%%%%%%%%%%%%%TABLE%%%%%%%%%%%%%%%%%%
\setlength{\tabcolsep}{8pt}
\begin{table}[!ht]
	\centering
	\renewcommand{\arraystretch}{1.2}
	\caption{Our final sample of galaxy clusters and groups, ordered by decreasing M$_{500}$. The mass is taken from the MCXC catalogue, which is based on the ROSAT X-ray luminosity \citep{2011A&A...534A.109P}. For 4C+55.16 (not present in MCXC) the mass is taken from \citet{2021ApJS..253...56Z}.}
	\label{tab:sample}
		\begin{tabular}{l|c|c|c}
			\hline
			Name & redshift & kpc/'' & M$_{500}$  \\
			 &  &  & [$10^{14}$M$_{\odot}$]\\
    
			\hline
			Abell~2390    & 0.230 & 3.67 & 8.95 \\
            
            \hline
			ClG~J1532+3021  & 0.362 & 5.04 & 8.16 \\
   
            \hline
			4C+55.16  & 0.242 & 3.82 & 7.40 \\
   
   			\hline
            Abell~478     & 0.088 & 1.65 & 6.42 \\

            \hline
			RX~J1447.4+0827  & 0.376 & 5.16 & 4.60 \\	

            \hline
			Abell~1664    & 0.128 & 2.28 & 4.06 \\	

            \hline
			RXC~J1558.3-1410  & 0.097 & 1.79 & 3.87 \\

            \hline
			ZwCl~8276 & 0.076 & 1.44 & 3.24 \\	

            \hline
			Abell~496     & 0.033 & 0.65 & 2.91 \\

            \hline
			ZwCl~235  & 0.083 & 1.56 & 2.80 \\	
   
            \hline
            Abell~2052    & 0.035 & 0.69 & 2.49 \\

            \hline
			Abell~3581    & 0.022 & 0.44 & 1.08 \\

            \hline
			NGC~6338  & 0.027 & 0.55 & 0.87 \\	

            \hline
			IC~1262   & 0.033 & 0.65 & 0.86 \\	

            \hline
			NGC~5098  & 0.036 & 0.72 & 0.52 \\		

            \hline
			NGC~5044  & 0.009 & 0.18 & 0.51 \\			
			
			\hline
		\end{tabular}
		
\end{table}
%%%%%%%%%%%%%%%%%%%%%%%%%%%%%%%%%%%%%%%

\section{Data and Methods} \label{sec:data}
In the following, we describe the procedure used to calibrate and analyze the VLBA and {\it Chandra} data, and to obtain the final sample shown in Tab. \ref{tab:sample}.
\subsection{VLBA radio data}\label{subsec:vlbadata}
\paragraph{Data reduction\,} \,The 5~GHz VLBA observations presented in \citet{2014PhDT.......338H,2015MNRAS.453.1223H,2015MNRAS.453.1201H} were performed in phase referencing mode (project codes BE056A, BE056B, BE056C, BE063B, BE063C, BE063E, BE063F, BE063H, BE063I, BE063J, BE063K). The average time spent on each target source was 20 minutes. The strategy we adopted to process the data is based on standard reduction techniques in \texttt{AIPS}. We first performed a correction of Earth orientation parameters and ionospheric delays (tasks \texttt{VLBAEOPS, VLBATECR}). Then, we solved for delays and amplitudes, by applying digital sampling corrections (\texttt{VLBACCOR}), removing instrumental delays on phases (\texttt{VLBAMPCL}), calibrating the bandpass (\texttt{BPASS}), and applying the corrections (\texttt{VLBAAMP}). To correct the time-dependent delays in phases, we fringe-fitted the data (\texttt{FRING}). Sources with a flux density larger than 20 mJy -- 30 mJy were self-fringed, while the others were referenced to the phase calibrator. Ultimately, we applied the calibration to the data and averaged the different IFs. Between two and ten cycles of self-calibration were attempted, and we found a significant improvement in sensitivity only for sources with a flux density larger than $\sim$50 mJy.
\paragraph{Morphological classification of the sources\,} \,\citet{2014PhDT.......338H} classified their radio sources as extended or unresolved based on the morphology of the detected emission or as undetected. We re-analyzed the calibrated data using the software \texttt{DIFMAP}, which can be used to image, characterize the morphology, and measure the flux density of a source by fitting its calibrated visibilities with different functions (an unresolved $\delta$-function, a circular gaussian, an elliptical gaussian, or a combination of these). Based on our analysis, we classified as undetected the targets for which no component with a flux density above $3\times\sigma$ could be fit to the visibilities; as unresolved the targets described with a single component smaller than the beam area; and as extended the targets described by least two components. 
\\We verified the presence of extensions in all of the sources classified as extended in the parent catalogue of \citet{2014PhDT.......338H}. Out of the sources classified as unresolved in \citet{2014PhDT.......338H}, we found evidence for extended jet-like emission in NGC~5098, ZwCl~8276, and RX~J1447.4+0828. Specifically, \citet{2014PhDT.......338H} reported the possible presence of extended emission in NGC~5098 (based on a visual inspection), which we confirm to be detected by fitting the visibilities in \texttt{DIFMAP} with an unresolved $\delta$-function (the core) and a gaussian (the extended jet). The extended components we detected in ZwCl~8276 and RXJ~1447+0828 are located along the beam major axis and, on the basis of a by-eye inspection of the images, the sources would be classified as unresolved; however, the best fit with \texttt{DIFMAP} is a combination of a central bright component and two more components on each side of the core in both clusters. 
\\Of the sources classified as undetected in the parent sample of \citet{2014PhDT.......338H}, our analysis revealed that Abell~795 is in fact detected (but unresolved) at $20\sigma$ (with $\sigma=50$~$\mu$Jy/beam), having a flux density of 1~mJy. The location of the detected radio emission matches the optical center of the BCG (RA, DEC = 9:24:05.3017, +14:10:21.524), while the phase center of the observation is offset by $\sim$5$''$ north-west, which may explain why it was missed in the analysis of \citet{2014PhDT.......338H}.  
\\ Overall, the fitting procedure in \texttt{DIFMAP} allowed us to confirm the presence of jets (at more than 5$\sigma$ significance) in 23 sources (out of the starting 45 BCGs, see Sect. \ref{sec:select}).

\subsection{Chandra X-ray data}
\begin{figure*}[!ht]
   \centering
   \includegraphics[width=0.9\linewidth]{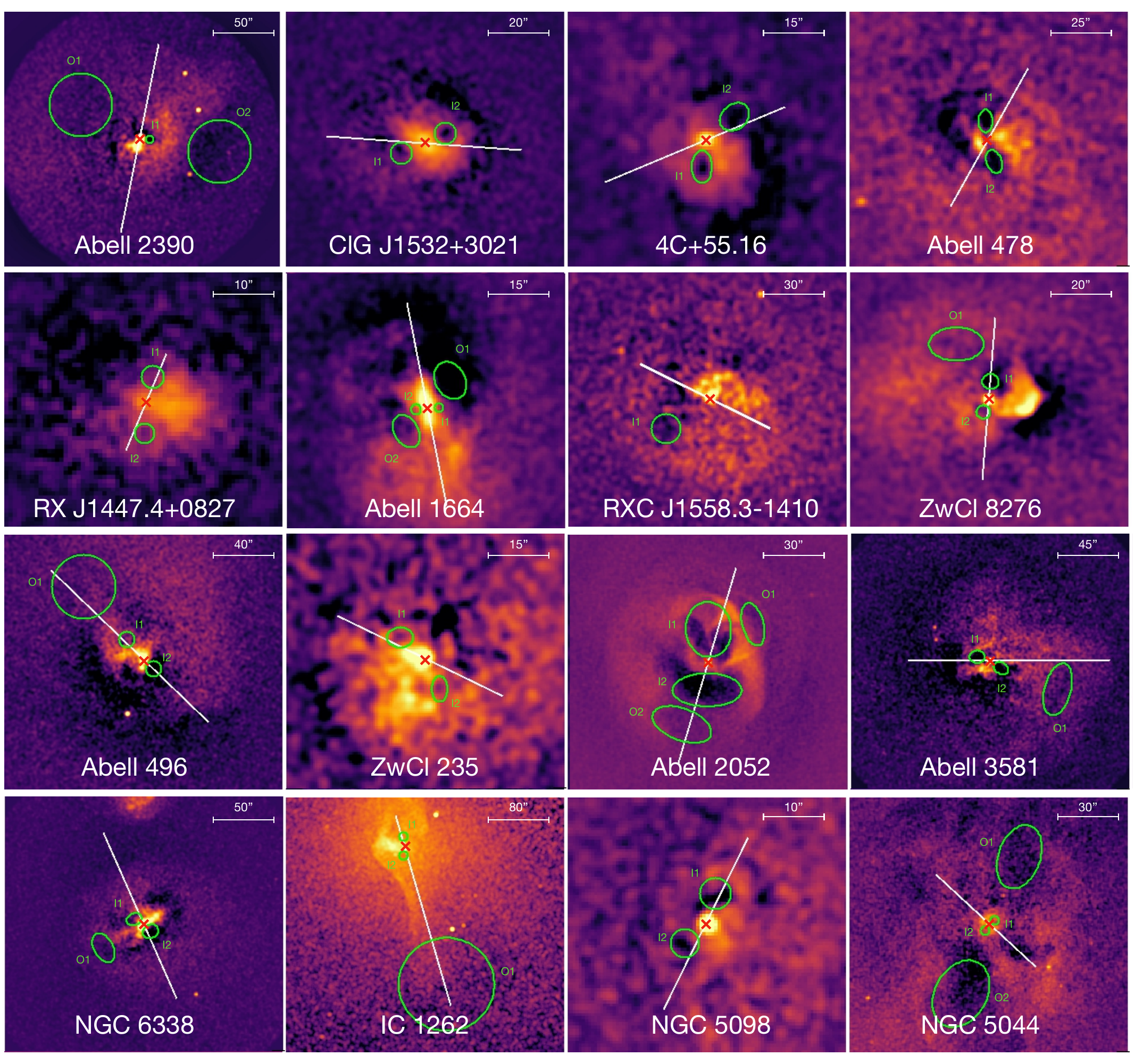}
      \caption{Residual $Chandra$ images of the systems in our sample. Overlaid in green are the X-ray cavities (see Tab. \ref{tab:cavpar} for details and literature references), while the white line shows the orientation of the radio jet in the plane of the sky, $\Phi_{\perp\text{LOS}}$. The red cross shows the position of the central AGN. See Sect. \ref{subsec:environment} for details.
              }
         \label{fig:sloshcompar}
\end{figure*}

\paragraph{Data reduction\,} \, We retrieved the data from the \textit{Chandra} archive (\url{cda.harvard.edu/chaser}) and reprocessed the observations using \texttt{CIAO-4.14}. Point sources were identified using the tool \texttt{wavdetect}, and masked during the analysis. For each target, we verified if the astrometry of the \textit{Chandra} observation needed further corrections beyond its nominal pointing accuracy (0.4$''$), and proceeded to update the coordinates using the \texttt{wcs\_match} and \texttt{wcs\_update} tools. We filtered the data from periods contaminated by background flares. The blank-sky event files have been selected as background file, and normalized by the 9-12 keV count-rate of the observation. 
\\To investigate, in detail, the morphology of the ICM, we produced images and corresponding background and exposure maps in the 0.5 -- 2 keV energy band. We also used the software \texttt{SHERPA} to fit the images with two-dimensional, elliptical double $\beta$ models and produce residual images, that can emphasize substructures in the ICM/IGrM (see Fig. \ref{fig:sloshcompar}).

%%%%%%%%%%%%%%%%%%%%% TABLE %%%%%%%%%%%%%%%%%%%%%%%
\setlength{\tabcolsep}{10pt}
\begin{table*}[!ht]
	\centering
	\renewcommand{\arraystretch}{0.8}
	\caption{Position and inclination angles of the radio jets from the VLBA data and position angle of the X-ray cavities from the $Chandra$ data. 
    (1) Cluster name. (2) Position angle in the plane perpendicular to the line of sight. (3) Jet/counter-jet ratio. (4) Inclination angle in the plane parallel to the line of sight. (5) Name of the cavity (I stands for inner, O stands for outer). (6) Cavity position angle. (7) Literature reference for the X-ray cavities. (8) Degree of misalignment between the jet direction and the X-ray cavity direction (measured as the difference between columns 2 and 6).}
	\label{tab:cavpar}
		\begin{tabular}{l|c|c|c|c|c|c|c}
			\hline
			Cluster & $\Phi_{\perp\text{LOS}}$ [$^{\circ}$] & R & $\Phi_{\parallel\text{LOS}}$  & Cavity & $\Theta_{\text{cav}}$ [$^{\circ}$] & Ref.  & $\Delta\Psi$ \\
    
			\hline
			Abell~2390       &  $100.9\pm28.5$  & 2.1 & $ 20.3 \pm 7.2 $   &   I1       &  5.2 $\pm$ 20.1  & 1, 2, 3 & $ 84.3  $  \\
                             &                  &     &                    &   O1       &152.4 $\pm$ 21.8  &         & $ 51.5  $  \\
                             &                  &     &                    &   O2       & -7.7 $\pm$ 22.4  &         & $ 71.4  $  \\
            \hline
			ClG~J1532+3021   &  $353.2\pm16.0$  & 3.6 & $ 35.4 \pm 11.3 $  &   I1       &203.6 $\pm$ 26.9  &  4      & $ 30.4  $  \\
                             &                  &     &                    &   I2       & 23.6 $\pm$ 26.9  &         & $ 30.4  $  \\
                            
            \hline
			4C+55.16         & $16.2\pm67.2$    &16.4 & $ 57.5 \pm 9.3 $   &   I1       &268.1 $\pm$ 29.7  &  5      & $ 71.9  $  \\
                             &                  &     &                    &   I2       & 36.7 $\pm$ 21.7  &         & $ 20.5  $  \\
			        
            \hline
			Abell~478        &  $241.3\pm25.4$  & 5.8 & $ 49.5 \pm 16.7 $  &   I1       & 99.0 $\pm$ 21.1  &  6, 7   & $ 37.7  $  \\		
                             &                  &     &                    &   I2       &206.5 $\pm$ 28.6  &         & $ 34.8  $  \\
            \hline
			RX~J1447.4+0827  & $68.1\pm15.5$    & 1.1 & $ 3.6 \pm 5.1 $    &   I1       &  80.4 $\pm$ 33.1 &  8      & $ 12.3  $  \\	
                             &                  &     &                    &   I2       & 265.6 $\pm$ 17.9 &         & $ 17.5  $  \\  
            \hline
			Abell~1664       &  $280.9\pm8.3$   & 6.0 & $ 49.9 \pm 16.3 $  &   I1       &  4.9 $\pm$ 24.5  &  9      & $ 84.0  $  \\	
                             &                  &     &                    &   I2       &184.9 $\pm$ 24.5  &         & $ 84.0  $  \\ 
                             &                  &     &                    &   O1       & 46.4 $\pm$ 30.6  &         & $ 54.5  $  \\	
                             &                  &     &                    &   O2       &226.4 $\pm$ 30.6  &         & $ 54.5  $  \\     

            \hline
			RXC~J1558.3-1410 &  $153.4\pm9.6$   & 4.2 & $ 39.8 \pm 12.8 $  &I1$^{\ast}$ &215.1 $\pm$ 19.7  & This work      & $ 61.7  $  \\
		
            \hline
			ZwCl~8276        & $265.8\pm60.3$   & 1.0 & $ 0.3 \pm 5.0 $    &   I1       & 77.9 $\pm$ 30.2  &  10, This work     & $ 7.9 $    \\	
                             &                  &     &                    &   I2       &257.9 $\pm$ 30.2  &         & $ 7.9 $    \\  
                             &                  &     &                    & O1$^{\ast}$&114.9 $\pm$ 23.9  &         & $ 29.1 $   \\
            \hline
			Abell~496        &  $136.9\pm15.6$  & 1.1 & $ 2.4 \pm 5.0 $    &  I1        &126.6 $\pm$ 17.8  &  11, 12 & $ 10.3  $  \\	
                             &                  &     &                    &   I2       &322.0 $\pm$ 33.1  &         & $ 5.1 $    \\  
                             &                  &     &                    &   O1       &128.9 $\pm$ 20.5  &         & $ 8.0 $    \\  
            \hline
			ZwCl~235         &  $153.4\pm38.9$  & 1.3 & $ 7.2 \pm 5.3 $    &  I1        &141.7 $\pm$ 17.3  &  13     & $ 11.7  $  \\	
                             &                  &     &                    &   I2       &291.3 $\pm$ 13.1  &         & $ 42.1  $  \\ 
                            
            \hline
			Abell~2052       &  $253.3\pm17.5$  & 1.1 & $ 3.6 \pm 5.1 $    &  I1        & 84.3 $\pm$ 49.5  &  14     & $ 11.0  $  \\	
                             &                  &     &                    &   I2       &271.6 $\pm$ 56.8  &         & $ 18.3  $  \\ 
                             &                  &     &                    &   O1       & 37.1 $\pm$ 20.8  &         & $ 36.2  $  \\	
                             &                  &     &                    &   O2       &248.2 $\pm$ 25.4  &         & $ 5.1 $    \\ 

            \hline
			Abell~3581       &  $189.7\pm59.8$  &10.2 & $ 54.1 \pm 12.4 $  &  I1        &163.3 $\pm$ 27.4  &  15, 16 & $ 26.4  $  \\	
                             &                  &     &                    &   I2       &325.9 $\pm$ 26.8  &         & $ 43.8  $  \\  
                             &                  &     &                    &   O1       &338.8 $\pm$ 21.2  &         & $ 30.9  $  \\
            \hline
			NGC~6338         &  $293.7\pm21.2$  & 3.3 & $ 32.9 \pm 10.9 $   &  I1        &137.9 $\pm$ 44.2  &  17, 18 & $ 24.2  $  \\	
                             &                  &     &                    &   I2       &317.9 $\pm$ 44.2  &         & $ 24.2  $  \\  
                             &                  &     &                    &   O1       &210.0 $\pm$ 20.2  &         & $ 83.7  $  \\

            \hline
			IC~1262          &   $284.2\pm15.6$ & 7.0 & $ 51.2 \pm 15.1 $  &  I1        & 85.3 $\pm$ 24.1  &  19     & $ 18.9$    \\	
                             &                  &     &                    &   I2       &265.3 $\pm$ 24.1  &         & $ 18.9$    \\ 
                             &                  &     &                    &   O1       &286.9 $\pm$ 25.1  &         & $ 2.7 $    \\	

            \hline
			NGC~5098         &  $244.7\pm21.2$  & 6.7 & $ 50.8 \pm 15.4 $  &  I1        & 72.1 $\pm$ 32.5  &  20     & 7.4        \\	
                             &                  &     &                    &   I2       &221.1 $\pm$ 30.3  &         & 23.6       \\    
            
            \hline
			NGC~5044         &  $140.3\pm19.4$  & 1.2 & $ 4.4 \pm 5.3 $    &  I1        & 68.7 $\pm$ 40.5  &  21, 22     & 71.6       \\
                             &                  &     &                    &   I2       &209.4 $\pm$ 35.9  &         & 69.1       \\ 
                             &                  &     &                    &   O1       & 64.5 $\pm$ 16.1  &         & 75.8       \\	
                             &                  &     &                    &   O2       &250.3 $\pm$ 21.9  &         & 70.0       \\
		
			\hline
		\end{tabular}
		\tablecomments{\scriptsize{[1] \cite{2011xru..conf..241L}; [2] \cite{2015ApSS.359...61S}; [3] \cite{2020MNRAS.496.2613B}; [4] \cite{2013ApJ...777..163H}; [5] \cite{2011MNRAS.415.3520H}; [6] \cite{2003ApJ...587..619S}; [7] \cite{2014ApJ...781....9G}; [8] \cite{2020AJ....160..103P}; [9] \cite{2019ApJ...875...65C}; [10] \cite{2013A&A...555A..93E}; [11] \cite{giacintucci2016}; [12] \cite{2002ASPC..262..377D}; [13] \cite{2023A&A...670A..23U}; [14] \cite{2011ApJ...737...99B}; [15] \cite{2005MNRAS.356..237J}; [16] \cite{2013MNRAS.435.1108C}; [17] \cite{2019MNRAS.488.2925O}; [18] \cite{2023ApJ...948..101S}; [19] \cite{2019ApJ...870...62P}; [20] \cite{2009ApJ...700.1404R}; [21] \cite{2014MNRAS.437..730O}; [22] \cite{2021ApJ...906...16S}.}\\
        $^{\ast}$: the X-ray cavity was identified in this work (see Appendix \ref{app:r1558} for RXC~J1558.3-1410 and Appendix \ref{app:comment} for the outer cavity in ZwCl~8276).
  }
\end{table*}

%%%%%%%%%%%%%%%%%%%%%%%%%%%%%%%%%%%%%%%%%%%%%%%%%%%

%%%%%%%%%%%%%%%%%%%%%%%%FIGURE%%%%%%%%%%%%%%%%%%%%%%%
\begin{figure*}[!ht]
   \centering
   \includegraphics[width=0.75\linewidth]{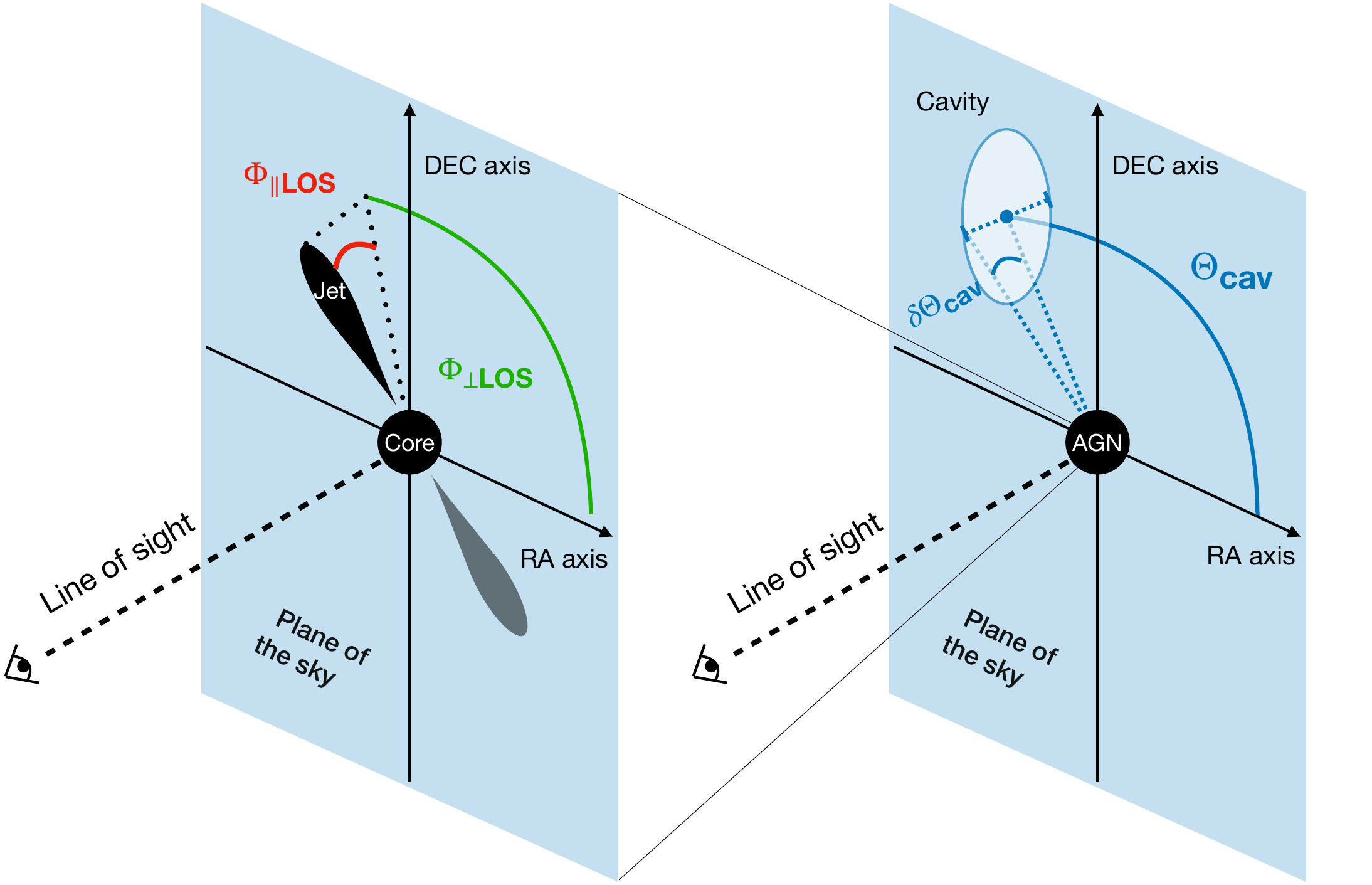}
      \caption{{\it Left:} schematic representation of a core$+$jet system, showing the jet position angle $\Phi_{\perp\text{LOS}}$ and the jet inclination angle $\Phi_{\parallel\text{LOS}}$. {\it Right:} schematic representation of an AGN $+$ cavity system, showing the bubble position angle in the plane of the sky, $\Theta_{\text{cav}}$, and the associated uncertainty, $\delta\Theta_{\text{cav}}$. 
              }
         \label{sketch-posangle}
\end{figure*}
%%%%%%%%%%%%%%%%%%%%%%%%%%%%%%%%%%%%%%%%%%%%%%%%%%%%%
%%%%%%%%%%%%%%%%%%%%%%%%FIGURE%%%%%%%%%%%%%%%%%%%%%%%
\begin{figure*}[!ht]
   \centering
   \includegraphics[width=0.9\linewidth]{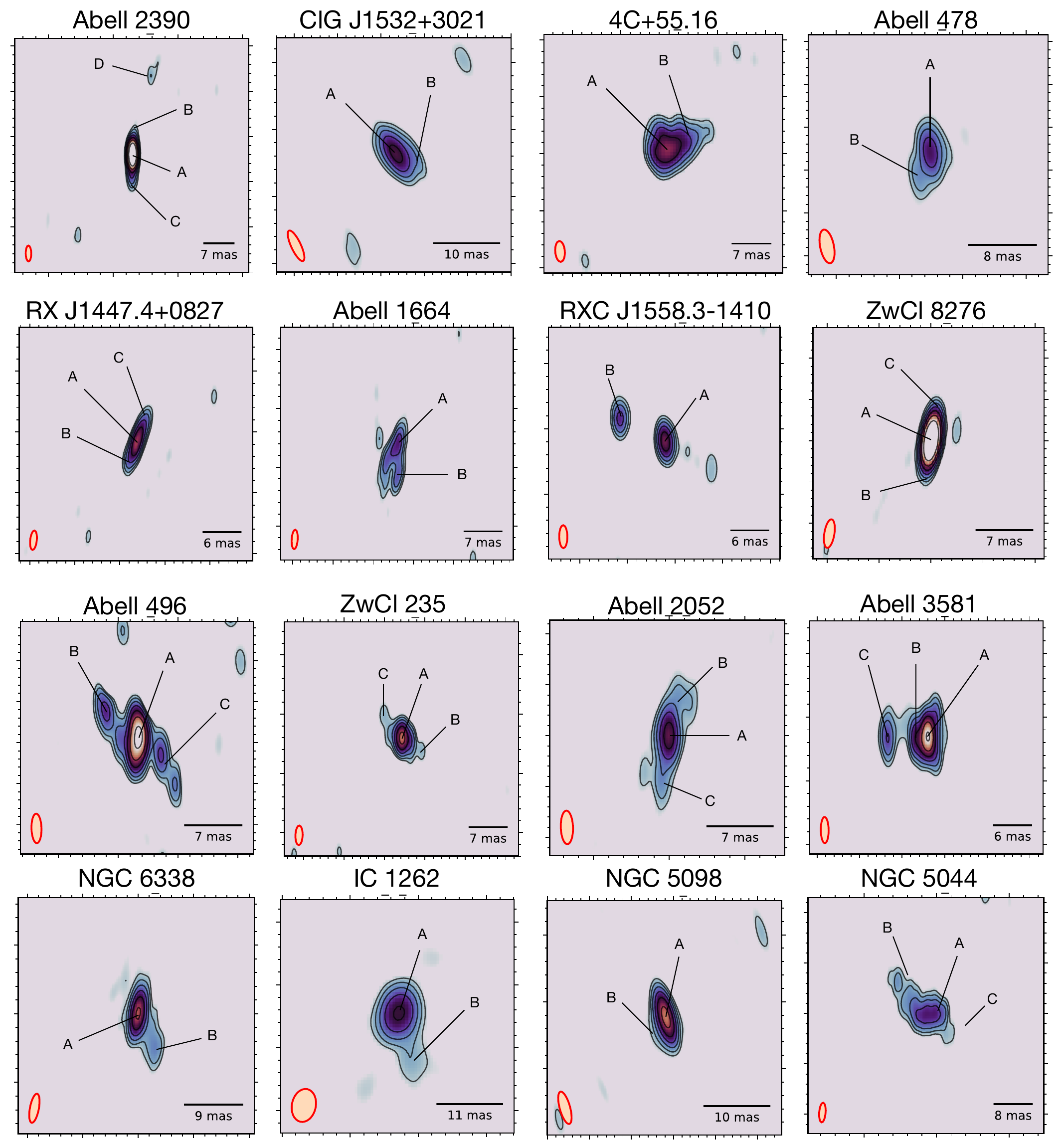}
      \caption{VLBA images at 5~GHz of the systems in our sample. Labels indicate the position of the components identified using \texttt{DIFMAP} (see Tab. \ref{tab:difmapcomp}). The contours overlaid in black start from $5\times\sigma$ (see the values in Tab. \ref{tab:difmapcomp}) and increase by a factor of 2. The beam is overlaid with an orange ellipse in the bottom left corner. See Sect. \ref{subsec:vlbadata} for details. 
              }
         \label{fig:vlbaima}
\end{figure*}
%%%%%%%%%%%%%%%%%%%%%%%%%%%%%%%%%%%%%%%%%%%%%%%%%%%%%
%%%%%%%%%%%%%%%%%%%%%%%%FIGURE%%%%%%%%%%%%%%%%%%%%%%%
%%%%%%%%%%%%%%%%%%%%%%%%%%%%%%%%%%%%%%%%%%%%%%%%%%%%%
\vspace{2pt}
\section{Measurement of position angles}\label{sec:measureangle}
In the following, we describe the methods we employed to measure the position angle of jets (on the parsec scale) and of X-ray cavities (on the kpc scale). A schematic representation is shown in Fig. \ref{sketch-posangle}.

\paragraph{Measuring the jet position angle in the plane of the sky\,}\, The \texttt{DIFMAP} analysis returned the integrated flux (F) and position relative to phase center in polar coordinates (r, $\theta$) of each component (see Appendix \ref{app:difmap}). To measure the position angle of the jet in the reference frame centered on the core, we adopted the following approach. We identified the core as the component with the highest radio flux density in the image (see Fig. \ref{fig:vlbaima}). The position angle $\Phi_{\perp\text{LOS}}$ in the plane of the sky of the extended jet-like component at ($r_{2}$, $\theta_{2}$) in the reference frame centered on the core at ($r_{1}$, $\theta_{1}$) is defined as: 
\begin{equation}
    \Phi_{\perp\text{LOS}} = \arctan\left(\frac{r_{2}\sin\theta_{2} - r_{1}\sin\theta_{1}}{r_{2}\cos\theta_{2} - r_{1}\cos\theta_{1}}\right),
\end{equation}
that is the angle between the line connecting ($r_{2}$, $\theta_{2}$) to ($r_{1}$, $\theta_{1}$) and the horizontal (right ascension) axis. 
If more than one jet is present (i.e. two sided sources), we considered the average between the angles of the two jets as the best estimate of the overall position angle. 
\\We note that the statistical errors on the parameters obtained from the \texttt{DIFMAP} fit may underestimate the real uncertainties (see e.g., \citealt{2018ApJ...862..139D}). As a more robust estimate of the position angle uncertainty, $\delta\Phi_{\perp\text{LOS}}$, we used the deconvolved angular width of the component (circle or ellipse) describing the jet emission at its distance from the core. In cases where the jet emission was unresolved by the \texttt{DIFMAP} fitting procedure (i.e. the component is a $\delta$-function), we considered the angular width of the beam at the distance of the jet from the core. In practice, given a jet-like component described by an ellipse with major semi-axis $a$ and minor semi-axis $b$, and located at a distance $R$ from the core, the uncertainty (in degrees) was determined as: 
\begin{equation}
    \delta\Phi_{\perp\text{LOS}} = \left(\frac{\sqrt{ab}}{R}\right)\times\frac{180}{\pi}\,\text{,}
\end{equation}
where $a$ and $b$, in the case of an unresolved component, are the beam major semi-axis and minor semi-axis, respectively. We list in Tab. \ref{tab:cavpar} our measurements of the jet position angle perpendicular to the line of sight for the systems in our sample. In Appendix \ref{app:altmethod} we show the comparison with two alternative methods that provided consistent results with those presented here. Furthemore, we show in Appendix \ref{subsec:otherfreq} that using archival VLBA data at different frequencies returns position angle measurements that are in good agreement with those derived here from the 5 GHz radio maps.

\paragraph{Measuring the jet inclination angle along the line-of-sight\,} \, As we are interested in the 3D geometry of the jets, we used the jet/counter-jet flux ratio to constrain the inclination angle of the jets with respect to the line of sight, which allows us to weight the importance of projection.
The jet/counter-jet flux ratio, $R_{j}$, is defined as:
\begin{equation}
    R_{j} = \frac{\text{F}_{\text{jet}}}{\text{F}_{\text{c-jet}}}
\end{equation}
where $\text{F}_{\text{jet}}$ and $\text{F}_{\text{c-jet}}$ are the integrated fluxes of the jet and of the counter-jet, respectively, measured in regions placed over each feature. For one-sided objects, the counter-jet flux was assumed to be lower than $5\times\sigma$. The inclination angle of the jet in a plane parallel to the line of sight, $\Phi_{\parallel\text{LOS}}$, can be computed using (e.g., \citealt{1979Natur.277..182S}): 
\begin{equation}
\label{eq:jcjratio}
    \cos\Phi_{\parallel\text{LOS}} = \frac{1}{\beta}\times\left(\frac{R_{j}^{\frac{1}{2+\alpha}}\,-\,1}{R_{j}^{\frac{1}{2+\alpha}}\,+\,1}\right),
\end{equation}
where $\alpha$ is the radio spectral index of the jets, assumed to be $\alpha = -0.6$. As the VLBA data probe the parsec scales, where the jets are still mildly relativistic, we assumed a Lorentz factor $0.6 \leq \beta \leq 1$ (see e.g., \citealt{2004ApJ...600..127G,2011IAUS..275..150G}). This returned a possible range of inclination angles along the line of sight. Thus, we used the median angle as our estimate of $\Phi_{\parallel\text{LOS}}$, and the range width as the statistical uncertainty. We also added in quadrature a systematic uncertainty of 5$^{\circ}$ (that is $R_{j}=1.2$, which could be intrinsic; see e.g., \citealt{1991ApJ...371..478M}).

\paragraph{Measuring the X-ray cavities position angle\,} \,  Out of the 23 objects where we could detect one or two sided jets from the VLBA data, there are literature indications for X-ray cavities in 15 objects. We inspected the counts and residual images of the other 8 objects, searching for $\sim20\%-30\%$ decrements in surface brightness at a signal-to-noise ratio greater than 3. We found one object where a clear cavity was visible, that is RXC~J1558.3-1410 (see Appendix \ref{app:r1558}). \citet{2013MNRAS.431.1638H} noted the presence of cavities in this cluster, but without detailing its/their position(s). Thus, we are unable to confirm that the structure we detect is coincident with the cavities they identified. 
We also found evidence of an external cavity in ZwCl 8276, besides the two inner bubbles noticed by \citet{2013A&A...555A..93E}. The existence of this external cavity, (shown in Fig. \ref{fig:sloshcompar}), is supported by the LOFAR detection of lobe-like radio emission matching the position of this outer bubble \citep{2020MNRAS.496.2613B}.
\\Overall, there are 16 systems in which the $Chandra$ data allows us to measure the position angle of at least one X-ray cavity. The following method was systematically adopted to measure the position angle of the cavities in the $Chandra$ images of the 16 systems. For each cluster or group, we measured the angle $\Theta_{\text{cav}}$ between the center of the ellipse describing a cavity and the RA axis with respect to the core of the central AGN, whose precise location is known from the VLBA data. The uncertainty in the position angle of the cavity, $\delta\Theta_{\text{cav}}$, is given by the width of the cavity in the direction perpendicular to the line connecting the AGN and the center of the cavity (for a schematic representation see Fig. \ref{sketch-posangle}, {\it right panel}). 
\\ We point out that we do not measure the inclination angle of cavities to the line of sight, as done for the radio jets. Relativistic arguments can constrain projection effects of radio jets on the pc-scale (see Eq. \ref{eq:jcjratio} and \citealt{1979Natur.277..182S}), whereas the depth of cavities in a cluster or group atmosphere is non-trivial to directly estimate from X-ray observations. Very few works have attempted to constrain cavity inclination angles, only in systems with remarkably deep cavities (as in e.g., RBS~797, \citealt{2011ApJ...732...71C}) and with several assumptions. Besides, the detectability of X-ray cavities is hampered when the structures lie at large angles (greater than 45$^{\circ}$) from the plane of the sky (see \citealt{2002A&A...384L..27E,2022A&A...665A..48M}), so that the majority of detected X-ray cavities in groups and clusters lie at less than 45$^{\circ}$ from the plane of the sky \citep{2009MNRAS.395.2210B}. Thus, we assume that the X-ray cavities in our sample are not far from the plane of the sky. While this represents a limitation of this work, we are confident that any tentative cavity inclination angle would likely be smaller than the associated uncertainty.  
\linebreak
\\ \indent Out of the 45/59 objects of \citet{2015MNRAS.453.1201H,2015MNRAS.453.1223H} with more than 10~ks of $Chandra$ data, the above procedures led us to consider only the 23/45 with one- or two-sided jets, and then the 16/23 with at least one X-ray cavity. Our final sample, thus comprising 16 objects, is listed in Tab. \ref{tab:sample}. The sample covers a mass range of $5\times10^{13}$ M$_{\odot}\leq M_{500} \leq 9\times10^{14}$ M$_{\odot}$, and is composed of 11 galaxy clusters and 5 galaxy groups. For each object we were able to measure the position and inclination angles of the jets ($\Phi_{\perp\text{LOS}}$, $\Phi_{\parallel\text{LOS}}$) and the position angle of the cavities ($\Theta_{\text{cav}}$). These values are reported in Tab. \ref{tab:cavpar}.

\section{Results} \label{sec:results}
%%%%%%%%%%%%%%%%%%%%%%%%FIGURE%%%%%%%%%%%%%%%%%%%%%%%
\begin{figure*}[!ht]
   \centering
   \includegraphics[width=0.95\linewidth]{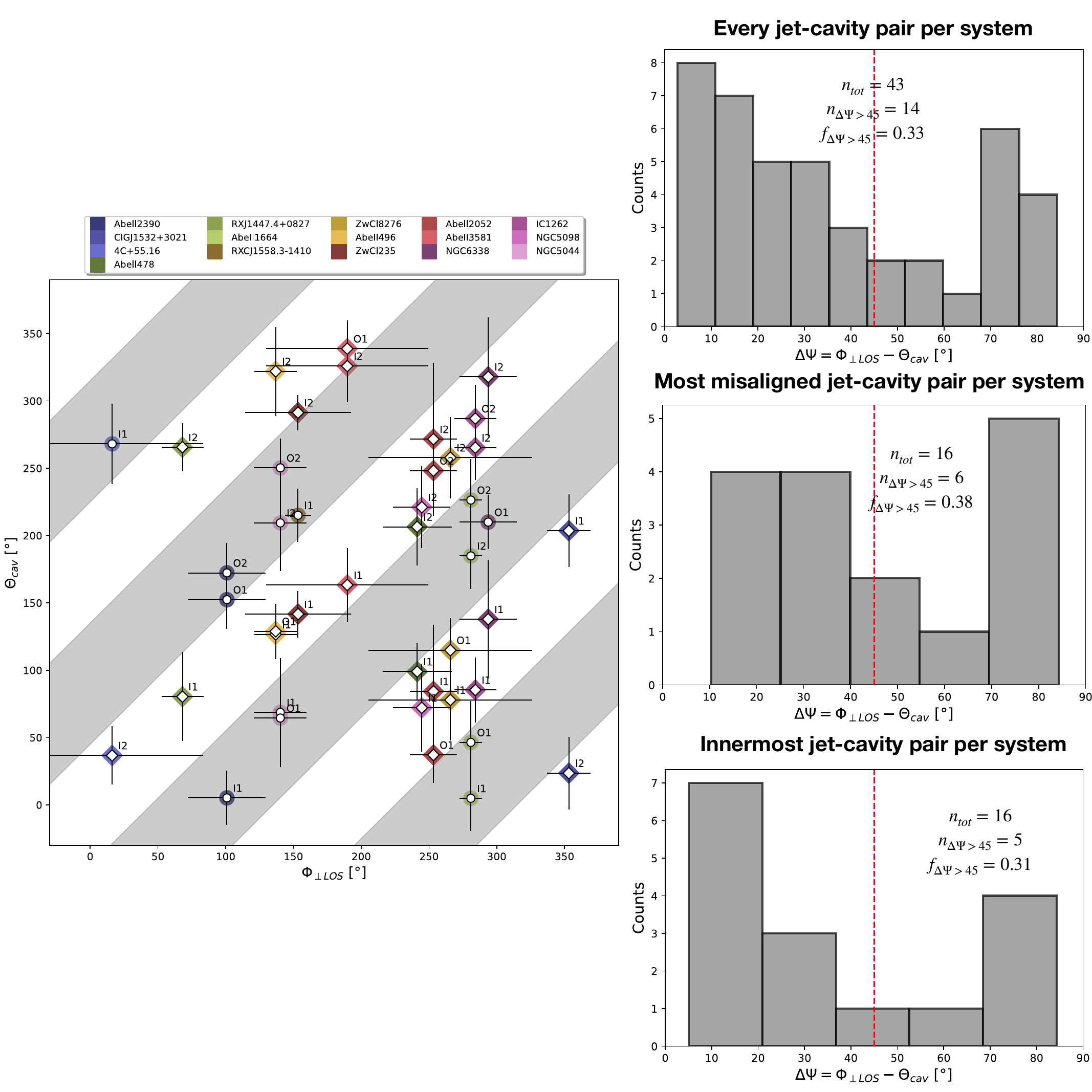}
      \caption{{\it Left:} Plot of jet position angle in the plane of the sky versus cavity position angle for the 16 systems in our sample. Each point is a jet -- cavity pair, meaning that systems with more than one X-ray cavity are represented by more than one point on the plot. The gray shaded areas represent spaces of the plot where the difference $\Delta\Psi = \Phi_{\perp\text{LOS}} - \Theta_{cav}$ is greater than 45$^{\circ}$. Jet -- cavity pairs falling inside these areas are represented with circles, while those falling outside these areas are represented with squares. The points are color-coded according to the cluster or group they belong to (see the top legend). Labels on the right of each point indicate the corresponding cavity (see Tab. \ref{tab:cavpar}). 
      \textit{Top right}: histogram of $\Delta\Psi$ for the 43 jet -- cavity pairs of the systems in our sample. 
      %The orange line is the fit with a Gaussian distribution to the systems with $\Delta\Psi\leq45^{\circ}$ (see Sect. \ref{subsec:likelihood} for details). 
      \textit{Center right}: histogram of the jet -- cavity pairs with the largest $\Delta\Psi$ for each systems.
      \textit{Bottom right}: histogram of the jet -- innermost cavity pair for each system.
      In each panel, the red dashed line shows the threshold of $45^{\circ}$ between aligned and misaligned jet -- cavity pairs.
              }
         \label{fig:compar-thetaphi}
\end{figure*}
%%%%%%%%%%%%%%%%%%%%%%%%%%%%%%%%%%%%%%%%%%%%%%%%%%%%%
\subsection{Jet -- cavity misalignments in the plane of the sky}\label{subsec:misal}
Focusing on the position angles in the plane of the sky of jets and X-ray cavities, we defined the quantity $\Delta\Psi = |\Phi_{\perp\text{LOS}} - \Theta_{cav}|$, which represents the degree of misalignment between the jet direction (measured on parsec scales) and the X-ray cavity direction (measured on kiloparsec scales). The values that $\Delta\Psi$ can assume range between 0$^{\circ}$ -- 90$^{\circ}$, with $\Delta\Psi = 0^{\circ}$ indicating a perfect alignment and $\Delta\Psi = 90^{\circ}$ indicating an orthogonal misalignment. 
For the sake of clarity, we note that for a cluster or a group with a number $N$ of X-ray cavities, there are $N$ pairs of ($\Phi_{\perp\text{LOS}}$, $\Theta_{cav}$) for which an estimate of $\Delta\Psi$ can be obtained. 
\\From now on, we arbitrarily classify as {\it aligned} the jet -- cavity pairs for which $\Delta\Psi < 45^{\circ}$, and as {\it misaligned} the jet -- cavity pairs for which $\Delta\Psi > 45^{\circ}$. In the last column of Tab. \ref{tab:cavpar} we list the values of $\Delta\Psi$ for the cavities of each cluster or group in our sample. Additionally, we show in Fig. \ref{fig:compar-thetaphi} a plot of $\Theta_{cav}$ versus $\Phi_{\perp\text{LOS}}$ for the jet -- cavity pairs in our sample. Given that the $180^{\circ}$ ambiguity in the definition of the position angle does not allow one to immediately evaluate $\Delta\Psi$ from this figure, we overplot as gray shaded areas the regions where $\Delta\Psi > 45^{\circ}$.
\\Among the 16 systems in our sample, there are 10 systems in which all the detected X-ray cavities are aligned with the radio jet seen at 5~GHz; these are ClG~J1532+3021, Abell~478, RX~J1447.4+0827, ZwCl~8276, Abell~496, ZwCl~235, Abell~2052, Abell~3581, IC~1262 and NGC~5098. An additional 2/16 objects, 4C+55.16 and NGC~6338, have at least one X-ray cavity that is aligned with the radio jet. In the remaining 4/16 objects, Abell~2390, Abell~1664, RXC~J1558.3-1410 and NGC~5044, all the detected X-ray cavities are found at more than 45$^{\circ}$ from the axis of the radio jet. We provide a short summary of the properties of each system in Appendix \ref{app:comment}.
\\ To further investigate the fraction of misaligned systems/X-ray cavities, we show in Fig. \ref{fig:compar-thetaphi} (top right) the histogram of $\Delta\Psi$ for the 43 jet -- cavity pairs of our sample. As expected from models of cavity formation, the majority of the X-ray cavities are aligned with the radio jet. However, as noted in the introduction, cases of spatial misalignment are known. Indeed, taking 45$^{\circ}$ as a threshold for a misaligned cavity-pair, there are 14/43 misaligned jet -- cavity pairs, that corresponds to a fraction of $33\%$. A similar argument can be made by considering, for each cluster or group, only the jet -- cavity pair with the largest $\Delta\Psi$. The corresponding histogram in Fig. \ref{fig:compar-thetaphi} (center right) shows that in 6/16 systems there is a cavity -- jet pair with $\Delta\Psi>45^{\circ}$, corresponding to a fraction of $38\%$. Moreover, restricting the histogram to the jet -- innermost cavity pair for each system (Fig. \ref{fig:compar-thetaphi}, bottom right), we find a fraction of $5/16 = 31\%$ of systems with $\Delta\Psi>45^{\circ}$. These results suggest that there is a 30\%-40\% chance to find a misalignment larger than $\Delta\Psi = 45^{\circ}$ when observing a cluster/group with a detected jet and at least one X-ray cavity.
\\We note that in the histograms of Fig. \ref{fig:compar-thetaphi} there is a peak at $\Delta\Psi\approx70^{\circ} - 80^{\circ}$, which comprises the cases where the jet and the cavity are nearly orthogonal. We discuss this feature and its interpretation in Sect. \ref{subsec:likelihood}.

\subsection{Jet inclination angle with respect to the line of sight}\label{subsec:anglelos}
As reported in Sect. \ref{sec:measureangle}, we constrained the jet inclination in a plane parallel to the line of sight using the jet -- counterjet flux density ratio (see Tab. \ref{tab:cavpar}). 
The angle $\Phi_{\parallel\text{LOS}}$ ranges between 0$^{\circ}$ (jets lying in the plane of the sky) to $\sim60^{\circ}$. There are 6/16 systems with $0^{\circ}\leq\Phi_{\parallel\text{LOS}}\leq10^{\circ}$ (RX~J1447.4+0827, ZwCl~8276, Abell~496, ZwCl~235, Abell~2052, and NGC~5044), 4/16 systems with $10^{\circ}\leq\Phi_{\parallel\text{LOS}}\leq45^{\circ}$  (Abell~2390, ClG~J1532+3021, RXC~J1558.3-1410, and NGC~6338), and 6/16 systems with $\Phi_{\parallel\text{LOS}}\geq45^{\circ}$ (4C+55.16, Abell~478, Abell~1664, Abell~3581, IC~1262, and NGC~5098). 
\\We notice that the lack of systems with jets nearly aligned along the line of sight is set by the combination (1) of our criterion of having at least one jet in the radio images (we are avoiding objects with $\Phi_{\parallel\text{LOS}}\approx90^{\circ}$); (2) of the assumptions on $\beta$ and $\alpha$ in Eq. \ref{eq:jcjratio} (that is, we do not consider superluminal motion, $\beta>1$, or jets with an inverted spectrum, $\alpha\geq0$); (3) of the limited dynamic range of the VLBA snapshot data ($\sim$20 min of time on source), thay may hide faint counterjets in one-sided sources; (4) of spherical geometry arguments: a given polar angle interval $\Delta\Phi$ covers more area on the sphere near the equator than near the poles, thus we expect only a small fraction of randomly pointed vectors to be pointed at us.
%%%%%%%%%%%%%%%%%%%%%%%%FIGURE%%%%%%%%%%%%%%%%%%%%%%%
\begin{figure*}[!ht]
   \centering
\includegraphics[width=\linewidth]{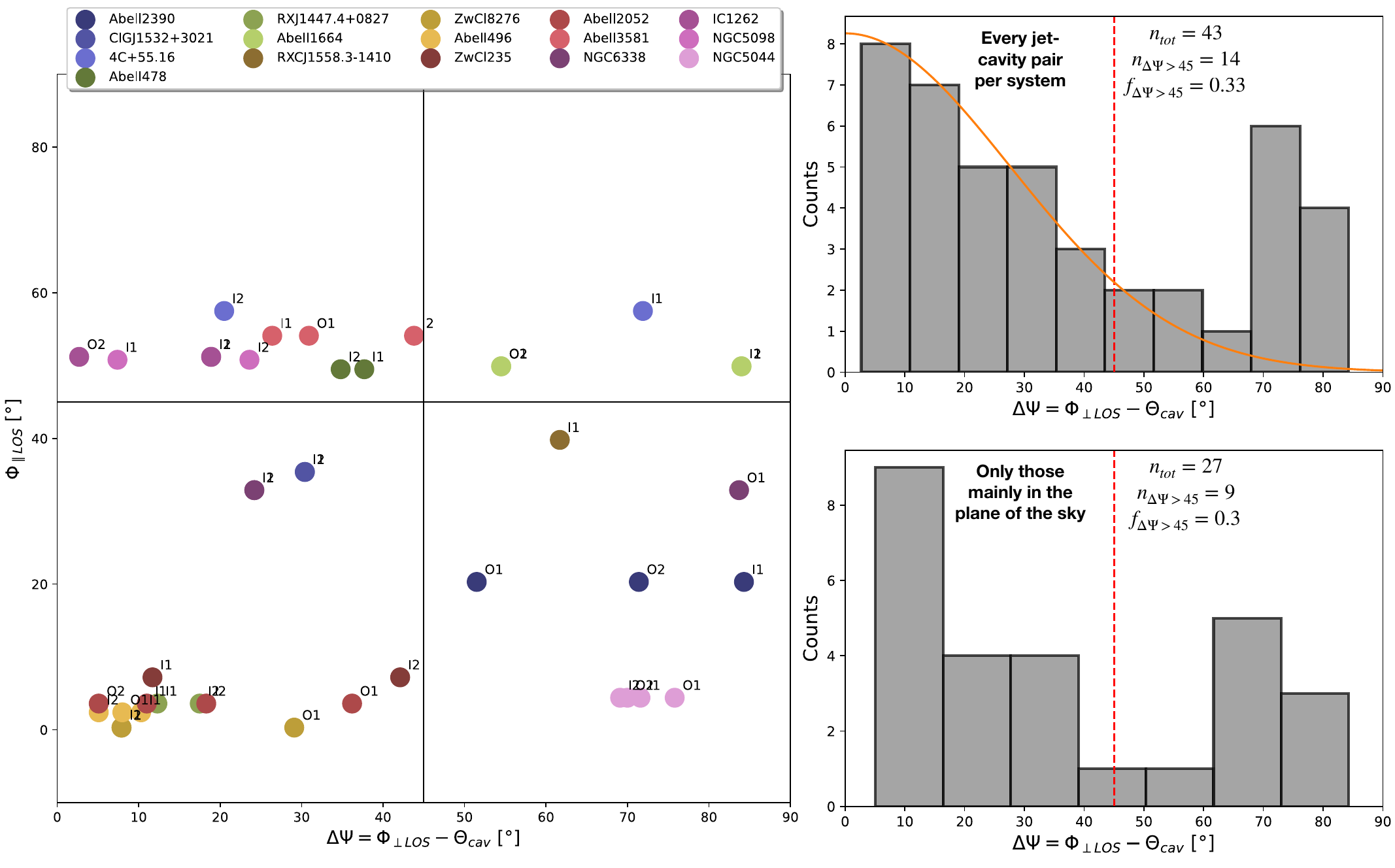}   
      \caption{{\it Left:} Diagnostic plot of misaligned jets and X-ray cavities. The jet inclination angle along the line of sight, $\Phi_{\parallel\text{LOS}}$, is plotted against the difference between the jet and the cavities position angles in the plane of the sky, $\Delta\Psi$. Points are color-coded according to the cluster or group they belong to (see the top legend). Labels on the right of each point indicate the corresponding cavity (see Tab. \ref{tab:cavpar}). \textit{Top right}: same as in Fig. \ref{fig:compar-thetaphi} (top right panel), but with the addition of the orange line, that is the fit with a Gaussian function to the systems with $\Delta\Psi\leq45^{\circ}$ (see Sect. \ref{subsec:likelihood} for details). \textit{Bottom right}: same as the top panel but including only the systems for which projection effects are negligible, that is $\Phi_{\parallel\text{LOS}}<45^{\circ}$. In both panels, the red dashed line shows the threshold of $45^{\circ}$ used to classify jet -- cavity pairs as aligned or misaligned.
              }
         \label{fig:scatter}
\end{figure*}
\section{Discussion} \label{sec:disc}
In the following, we test different scenarios to explain the alignments and misalignments between jets and X-ray cavities in our sample. After assessing the relative importance of external factors, that is projection effects and environmental disturbances, we turn to the discussion of effects related to the central engine. Specifically, we show how the distribution of $\Delta\Psi$ for the clusters and groups in our sample can provide insights on the likelihood of jet reorientation, and we consider the timescales of such mechanisms. 

\subsection{Projection effects} \label{subsec:projection}
As noted in Sect. \ref{subsec:misal}, there are systems in our sample where some X-ray cavities are misaligned by more than 45$^{\circ}$ from the radio jet of the AGN. However, there is the possibility that the misalignment is only apparent. Indeed, projection effects may cause a large apparent offset, even though the actual 3D misalignment between jets and cavity directions is small. Using the information obtained from the analysis of the jet direction with respect to the line of sight (Sect. \ref{sec:measureangle} and \ref{subsec:anglelos}) we can systematically reconstruct the 3D geometry of jets and verify whether projection effects may reconcile the largest misalignments we found. 
\\ We show in Fig. \ref{fig:scatter} (left panel) the jet inclination along the line of sight, $\Phi_{\parallel\text{LOS}}$, plotted against the difference between the jet and the cavity position angles in the plane of the sky, $\Delta\Psi$. This figure can be used as a diagnostic plot for the degree of misalignment of each jet -- cavity pair in the systems of our sample. As done before, we use $\Phi_{\parallel\text{LOS}}=45^{\circ}$ and $\Delta\Psi=45^{\circ}$ as thresholds between jets in the plane of the sky or along the line of sight and between aligned or misaligned jet -- cavity pairs, respectively. We can thus divide the 43 jet -- cavity pairs between the four quadrants of the plot, obtaining the following results: \\
\textbullet \,{\it Bottom-left quadrant:} aligned jet -- cavity pairs which are mainly lying in the plane of the sky. \\
\textbullet \,{\it Upper-left quadrant:} aligned jet -- cavity pairs where the jet is primarily oriented towards us. \\
\textbullet \,{\it Upper-right quadrant:} misaligned jet -- cavity pairs where the jet is primarily oriented towards us.\\
\textbullet \,{\it Bottom-right quadrant:} misaligned jet -- cavity where the jet is mainly lying in the plane of the sky.\\
Overall, the (mis)alignments of jet -- cavity pairs that are located in the two bottom quadrants, where projection effects are relatively negligible, are robustly confirmed even in 3D. By restricting the analysis to these jet -- cavity pairs (a total of 27 pairs), we still find that a fraction of $9/27 = 30\%$ is misaligned by more than 45$^{\circ}$, up to nearly orthogonal misalignments. Such fraction of misaligned jets in the plane of the sky is clearly independent of the inclination angle along the line of sight. Therefore, projection effects may account for an apparently large $\Delta\Psi$ only in a fraction ($\sim$35\%) of the misaligned jet -- cavity pairs in our sample (as in Abell~1664 or for the cavity I1 in 4C+55.16).

\subsection{The environment} \label{subsec:environment}
In this subsection we consider whether environmental effects may favor misalignments between successive AGN outbursts. First, we note that over the mass range covered by our sample ($5\times10^{13}$ M$_{\odot}\leq M_{500} \leq 9\times10^{14}$ M$_{\odot}$) there is no clear evidence for the degree of misalignment of the different jet -- cavity pairs being related to the mass of the host halo. Based on the 16 systems in our sample, dramatic changes in jet direction can be found both in massive galaxy clusters (such as Abell~2390, with $\Delta\Psi_{max} = 84.3^{\circ}$) and in galaxy groups (such as NGC~6338, with $\Delta\Psi_{max} = 83.7^{\circ}$, or NGC~5044, with $\Delta\Psi_{max} = 75.8^{\circ}$).
\\A potentially important question is whether gravitational perturbances that trigger large scale gas motions could play a major role. Specifically, cavities pointing to a different direction than the jets may have been shifted from their original axis by the gas motions. 
\\If gas bulk motions (such as sloshing) were responsible for the jet -- cavity misalignments, we should expect misaligned pairs to be more likely found in disturbed clusters and groups. Typical evidence of sloshing include a spiral morphology of the ICM/IGrM, and/or the presence of one or more cold fronts. We checked in the literature if the objects in our sample are reported to be in a sloshing state. Interestingly, all the systems are dynamically disturbed, with all but one (NGC~6338, which is merging with another group) showing evidence of sloshing. The relevant references are given in Tab. \ref{tab:cavpar}. Even though no dedicated X-ray studies are available for RXC~J1558.3-1410, we remark that the $Chandra$ image reveals a bright spiral residual feature in the ICM (see Fig. \ref{fig:r1558-appendix} in Appendix \ref{app:r1558}), which indicates that sloshing is likely present also in this system. 
\\Nice comparisons are offered by Abell~496, Abell~3581, Abell~1664 and NGC~6338 (see Fig. \ref{fig:sloshcompar}), which are all strongly disturbed systems. In Abell~496 and Abell~3581 the different generations of X-ray cavities are nearly aligned with each other and with the radio jet. Conversely, in Abell~1664 and NGC~6338 the different generations of X-ray cavities are largely misaligned by more than $45^{\circ}$ from each other and from the radio jet. 
Even though we cannot exclude that in some systems the present position of the X-ray cavities may have been influenced by the large scale motions, we argue that the environment does not play the major role, as gas dynamical perturbations are present in the whole sample. Sloshing may become a dominant mechanism for the dynamics of X-ray cavities at later times (as for the oldest, detached radio lobe of NGC~5044, see \citealt{2014MNRAS.437..730O}), since sloshing timescales (around $10^{8} - 10^{9}$~yr, e.g., \citealt{2017ApJ...851...69S}) are usually an order of magnitude larger than the age of X-ray cavities (around $10^{6} - 10^{7}$~yr, e.g., \citealt{2004ApJ...607..800B}; see also Sect. \ref{subsec:timescales}).  
%%%%%%%%%%%%%%%%%%%%%%%%%%%%%%%%%%%%%%%%%%%%%%%%%%%%%
\subsection{The likelihood of jet reorientation}\label{subsec:likelihood}
The above results argue against projections or environmental effects, alone, being able to explain the jet -- cavity misalignments.  All in all, it seems more likely that the large changes in direction are related to actual reorientation of the SMBH-driven jets. We now turn to the questions of how likely it is for jets to experience reorientation, and whether there is a favored angle. 
\\Our analysis of Sect. \ref{subsec:misal} unveiled that $\sim30\%-40\%$ of the jet -- cavity pairs in our sample have a $\Delta\Psi$ larger than 45$^{\circ}$. 
Furthermore, when excluding systems that may be biased by projection effects (see Sect. \ref{subsec:projection}), we find that $30\%$ of jet -- cavity pairs are misaligned by an angle (in the plane of the sky) ranging between 45$^{\circ}$ and 90$^{\circ}$. This suggests that a change in the jet position angle of AGN may be a relatively common mechanism, as roughly $1/3$ of the X-ray cavities in our sample are more than 45$^{\circ}$ away from the currently driven jets. 
\\A peculiar feature in the distribution of $\Delta\Psi$ is the peak around $\approx75^{\circ}$. We note that the peak is resilient to the systems aligned along the line of sight being removed from the histogram (see Fig. \ref{fig:scatter}, bottom right panel). 
If no changes in time of the position angle of the jets occurred, one might expect that cavities and corresponding jets should always be perfectly aligned, i.e. $\Delta\Psi = 0^{\circ}$. This is clearly an idealized case, as there are several effects that contribute to widening the distribution around zero. These can be both intrinsic to the physics of jets and X-ray cavities (such as slow precession of the jet axis over time, cavities expanding asymmetrically, or buoyant forces governing the motion of bubbles at later times\footnote{In case of a non-spherical potential well, a buoyant bubble will preferentially rise in the direction of the shallower potential (that is the minor axis of the dominant galaxy, in case of an elliptical morphology). However, the majority of radio galaxies already have their jets oriented within $\sim30^{\circ}$ of the optical minor axis of their host \citep{1979ApJ...231L...7P,2009MNRAS.399.1888B,2019arXiv190809989V}, which limits the magnitude of misalignments induced by an asymmetric potential.}) and related to observational biases (cavities not being perfect ellipses, resolution of radio and X-ray observations, projection effects). All of these likely result in a broadening of the distribution to a nearly Gaussian shape. 
This is indeed the case for the jet -- cavity pairs with $\Delta\Psi\leq45^{\circ}$: fitting a Gaussian function to this side of the distribution we find a mean $\mu = 0.0^{\circ}\pm1.2^{\circ}$ with standard deviation $\sigma = 27.6^{\circ}\pm1.6^{\circ}$ (see Fig. \ref{fig:scatter}, top right panel). This scatter of $\sim30^{\circ}$ expresses the above widening effects.  
\\ The effect of jet reorientation on the histogram depends on whether there is a favored angle by which jets are reoriented. If jet reorientation caused a random change in the position angle of the jets, one would expect a uniform distribution of misaligned jets and X-ray cavities combined with the Gaussian-like distribution at $\Delta\Psi<45^{\circ}$. However, the presence of the peak at $\Delta\Psi\sim75^{\circ}$ in Fig. \ref{fig:scatter} (right panels) may suggest that when a change in jet orientation between successive AGN outbursts occurs, the change in position angle is more likely to be large, rather than randomly distributed between $0^{\circ} - 90^{\circ}$. That is, we may cautiously propose that the mechanism of jet re-orientation favors large changes in jet position angle, causing nearly orthogonal misalignments between the successive AGN outbursts.
\\We note that these conclusions would strongly benefit from increasing the sample size to include a larger number of objects. We can estimate the robustness of the above result for our sample of 16 clusters with a total of 43 X-ray cavities. We performed $10^{3}$ realizations of drawing 43 values between $0^{\circ} - 90^{\circ}$ from the combination of a Gaussian (with $\mu$, $\sigma$ set to those derived above) and a uniform distribution. The relative number of points drawn from the Gaussian was set to 29/43 (the jet -- cavity pairs with $\Delta\Psi\leq45^{\circ}$), while the remaining 14/43 values (the jet -- cavity pairs with $\Delta\Psi\geq45^{\circ}$) were drawn from the uniform component. We find that the probability of having $n_{70^{\circ}-90^{\circ}}/n_{45^{\circ}-70^{\circ}} > 2$ (as observed) is of 1.6\%, thus we can reject the uniform distribution at $\sim2.5\sigma$ confidence. 
\\We also note that this is not the first time that a secondary peak at large angles is found when studying small and large scales misalignments. Earlier works found a  secondary peak at around $90^{\circ}$ \citep{Pearson1988,Wehrle1996,Cao2000}, more recently confirmed in \citet{Kharb2010}. These works were based on radio data (misalignments were evaluated through VLA and VLBA comparisons), and the sources of those samples were mainly blazars or quasars, for which projection effects are certainly not negligible. In our study we find a very similar result (a peak at $\Delta\Psi\sim70^{\circ} - 80^{\circ}$) using the X-ray counterparts of large scale radio lobes (the cavities), and, most importantly, even when restricting the analysis to sources where projection effects are negligible (see Fig. \ref{fig:scatter}, bottom right panel). Therefore, while limited by the statistics of our sample, we confirm the earlier claims of a bimodality in the distribution of misalignments. In turn, this supports the idea that the mechanism behind the changes in jet orientation favors large misalignments over small ones.

\subsection{The timescales and mechanism of jet reorientation} \label{subsec:timescales}
\begin{figure*}[ht!]
 \includegraphics[width=0.5\textwidth,trim=1.0cm 1.0cm 1.0cm 1.0cm]{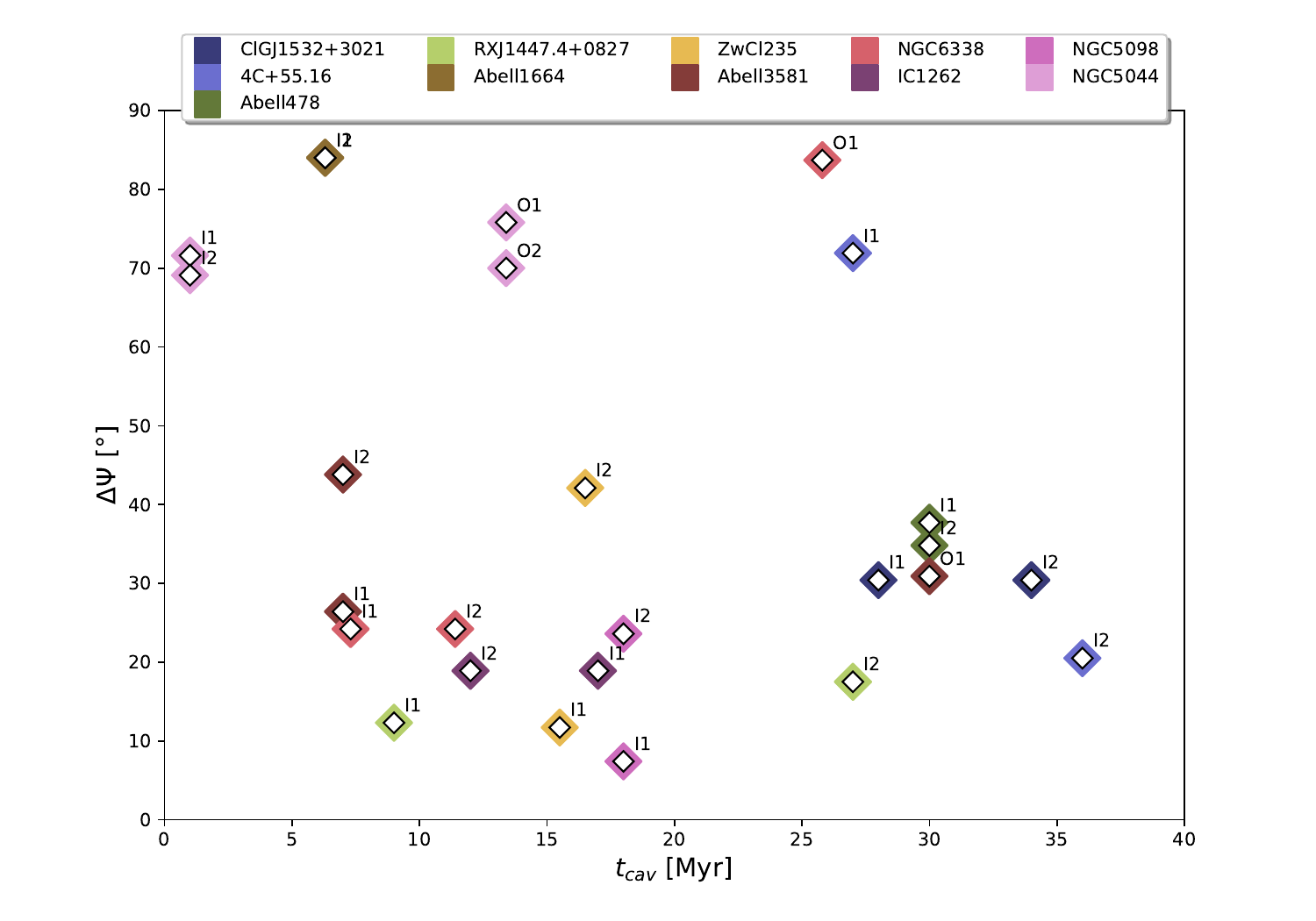}
    \includegraphics[width=0.5\textwidth,trim=1.0cm 1.0cm 1.0cm 1.0cm]{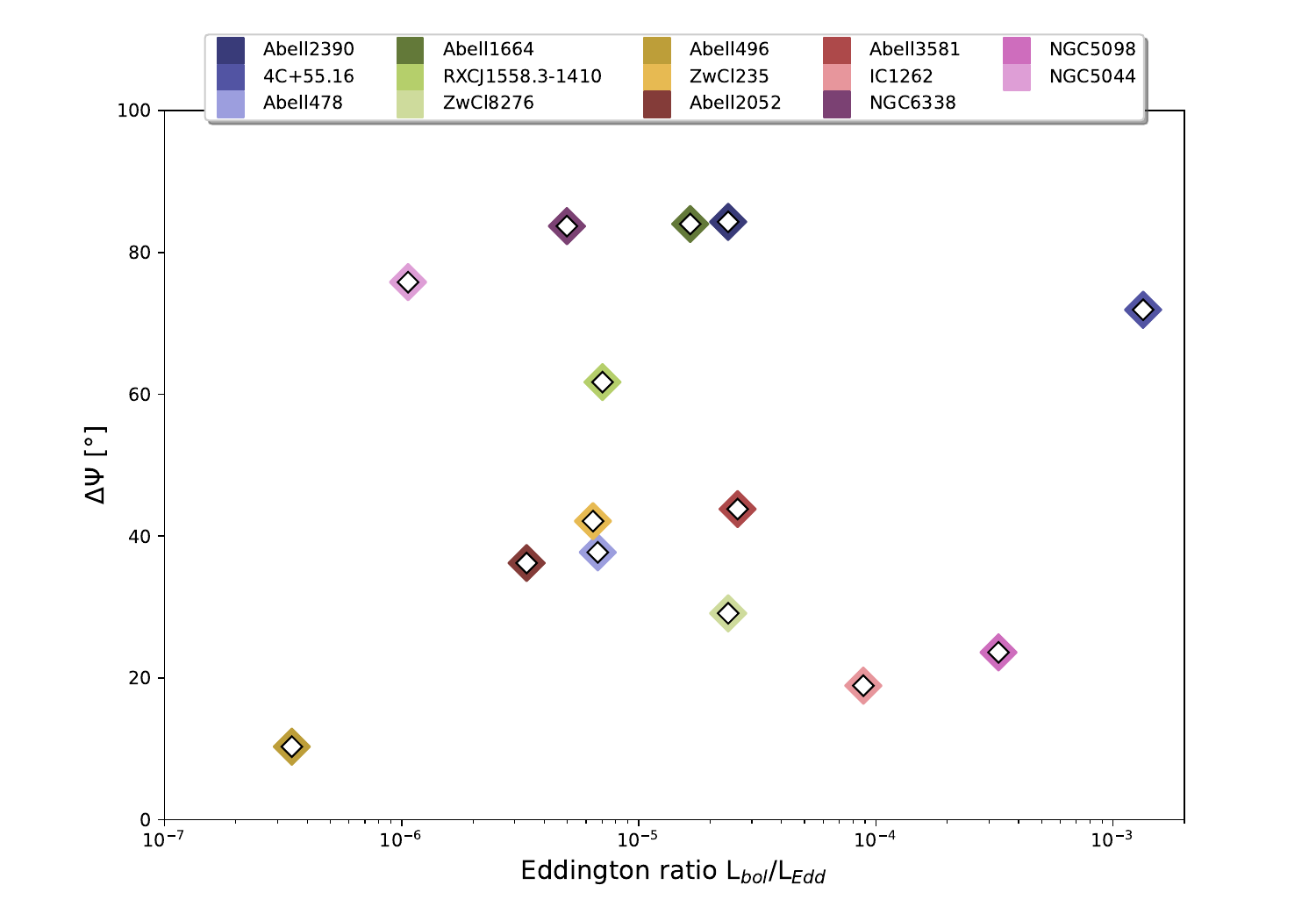}
  \caption{{\it Left panel:} Jet -- cavity position angle difference $\Delta\Psi$ versus cavity age $t_{cav}$ for the cavities in the systems of Tab. \ref{tab:cavpar} with an available estimate of the age from the literature. {\it Right panel:} Largest jet -- cavity position angle difference $\Delta\Psi$ for each system in our sample versus the Eddington ratio of the central AGN (see Sect. \ref{subsec:timescales} for details). In the right panel, only the 14/16 systems with available measurements of $L_{bol}$ and $L_{Edd}$ are plotted. }
  \label{fig:tcavcompar}
\end{figure*}

The observed misalignments raise the question of how quickly jets can change propagation direction over time. To investigate this point, we searched in the literature references listed in Tab. \ref{tab:cavpar} which X-ray cavities in our sample had an estimate of their ages (24 out of 43 X-ray cavities). We show in Fig. \ref{fig:tcavcompar} (left panel) a plot of $\Delta\Psi$ versus the age of each X-ray cavity (obtained through different methods, which typically agree within a factor of two; see \citealt{2004ApJ...607..800B}). No clear trend is visible in this figure. This may indicate that larger changes in jet axis orientation do not always occur on longer timescales. By focusing on the systems with $\Delta\Psi>45^{\circ}$, we see that the young X-ray cavities in NGC~5044 and in Abell~1664, likely formed only a few Myrs ago, are already misaligned by a large angle from the radio jet currently being driven by the central engine. The case of NGC~5044 is remarkable, with the 1 Myr old inner X-ray cavities being at $\sim70^{\circ}$ from the jet seen in VLBA images. \citet{2021ApJ...906...16S} discussed many possibilities for such misalignment, ruling out binary black holes and precession scenarios. 
%%%%%%%%%%%%%%%%%%%%%%%%FIGURE%%%%%%%%%%%%%%%%%%%%%%%
%%%%%%%%%%%%%%%%%%%%%%%%%%%%%%%%%%%%%%%%%%%%%%%%%%%%%
\\ In the case that the jet's direction of propagation is linked to the spin of the rotating SMBH only, dramatically altering the SMBH rotation axis on short timescales is non-trivial. To address this issue, the case has been made that the jet direction could be influenced by the accreting material. We point out that, in the following discussion, we do not assume a one-to-one relation between an accretion episode and a jet episode. Rather, during continuous gas accretion, the variations of the accretion rate, the magnetic field, or the spin of the SMBH may drive the formation of a relativistic jet (e.g., \citealt{blandford1977,meier2001,wu2008,cemeljic2022}).
\linebreak
\\\citet{2013ApJ...768...11B} argued that infalling clouds may form a thin accretion disk misaligned from the spin axis of the SMBH. In this case, the orientation of the disk would influence the direction of the emerging jet. Thin accretion disks are usually associated with radiatively efficient AGN with high accretion rates, such as quasars and Seyferts. By contrast, radiatively inefficient radio galaxies, which are common at the center of galaxy groups and clusters, are more likely powered by geometrically thick accretion flows with a low accretion rate \citep{ghisellini2001dividing}. 
We also note that our analysis in Sect. \ref{subsec:likelihood} suggests that the change in jet position angle over time is more likely to be nearly orthogonal, rather than completely random between $0^{\circ} - 90^{\circ}$. This is reminiscent of X-shaped radio galaxies, where the Bardeen -- Petterson mechanism \citep{bardeen1975} may cause orthogonal reorientation of the jets \citep{2004MNRAS.347.1357L}. However, X-shaped radio galaxies are preferentially high accretion rate AGN (e.g., \citealt{machalski2016,garofalo2020,giri2022}). 
\\ The question thus arises whether the systems with reoriented jets can have larger accretion rates, thus temporarily allowing the disk to influence the spin of the SMBH. To investigate this point, we derived the Eddington ratio of the systems in our sample. The Eddington ratio is defined as: 
\begin{equation}
\label{eq:eddratio}
    f_{\text{Edd}} = \frac{L_{bol}}{L_{Edd}},
\end{equation}
\noindent where $L_{bol}$ is the AGN bolometric luminosity, and $L_{Edd}$ is the Eddington luminosity. The latter is defined as:
\begin{equation}
\label{eq:eddlumin}
    L_{Edd} = 1.3\times10^{38}\,M[\text{M}_{\odot}]\,\,\text{erg/s},
\end{equation}
where $M$ is the mass of the SMBH (see \citealt{czerny2016} for a review). The $L_{bol}$ and $L_{Edd}$ for the majority of the systems in our sample were retrieved from \citet{mezcua2018}, who used the 2 -- 10 keV X-ray nuclear luminosity and a bolometric correction factor of 20 to estimate $L_{bol}$. For NGC~6338, NGC~5044, and 4C+55.16 the bolometric and Eddington luminosities were derived from the X-ray nuclear luminosity and black hole mass reported in \citet{2013MNRAS.432..530R}, respectively, adopting the same bolometric conversion factor of \citet{mezcua2018}. A similar method has been used to derive $L_{bol}$, $L_{Edd}$ from the literature references of NGC~5098 \citep{2009ApJ...700.1404R,2021ApJ...914..121A} and IC~1262 \citep{2020ApJ...900..124B,2021ApJ...914..121A}. No useful data is available for ClG~J1532+3021 and RX~J1447.4+0827. 
\\Fig. \ref{fig:tcavcompar} (right panel) shows the largest degree of misalignment of each cluster or group in our sample
versus the Eddington ratio. We find low levels of accretion, $f_{Edd}\leq10^{-3}$, for all the AGN, with an average of $\overline{f}_{Edd} = 1.5\times10^{-5}$. For comparison, the isolated X-shaped radio galaxies in \citet{2019ApJ...887..266J} have $\overline{f}_{Edd} = 4\times10^{-3}$. Additionally, the systems in our sample with and without reoriented jets have similar accretion efficiencies. 
\\Likely, the Eddington ratios we measure represent the {\it current} accretion efficiencies. Since the accretion rate can flicker over time (e.g., \citealt{schawinski2015,king2015}), we may hypothesize that the AGN with reoriented jets have been highly accreting in the past, thus allowing the Bardeen-Petterson effect to trigger the reorientation of the jets. While we cannot verify this hypothesis, the above result suggests that any high-accretion efficiency phase must have been short enough to allow the AGN to quickly transition back to a low-accretion efficiency phase (the one observed at present).
\\In this context, theoretical and computational works have argued that the radiatively inefficient, geometrically thick disk (ADAF-like, \citealt{1995ApJ...452..710N}) in the core of cluster and group-central galaxies may be fed by individually infalling molecular clouds, whose angular momentum varies over time (e.g., \citealt{2017MNRAS.466..677G,2022Univ....8..483S}). These parcels of cold gas may temporarily boost the accretion rate (e.g., \citealt{Gaspari2015}).
\\ For completeness, \citet{2022ApJ...936L...5L} showed how during the first phases of AGN activity the jets can naturally experience intermittency and fast (down to $10^{-2}$~Myr) changes in orientation, even at low accretion efficiencies. 
However, the timescales shown in Fig. \ref{fig:tcavcompar} (left panel) do not reflect a rapid flickering of jet direction in the first $\sim1$~Myr of AGN activity, before stabilizing to a collimated and stable structure. Indeed, in NGC~5044 there are at least two aligned outbursts (with an age of $\sim$13~Myr) that preceded the youngest, misaligned one (the one seen on VLBA scales), which indicates that flickering would have had to last at least an order of magnitude longer than the timescales explored in \citet{2022ApJ...936L...5L}. The same holds e.g., for NGC~6338, and likely for Abell~2390 (note that although there are no estimates of the X-ray cavity ages in this system, the 100 kpc -- scale bubbles O1 and O2 are plausibly older by $10 - 100$ Myr than the current radio jet - see also \citealt{2023MNRAS.522.1521A}).
\\ Therefore, if jet reorientation in cluster and group-central galaxies is driven by a temporary phase of high accretion efficiency, this phase must be short lived. \citet{2013ApJ...768...11B} and \citet{garofalo2020} estimated that the quasar-like phase should last a few Myrs to allow any appreciable changes in the jets direction. This is a reasonable timescale for most of the AGN in our sample, for which the cavity ages shown in Fig. \ref{fig:tcavcompar} (left panel) are larger than $\sim5-10$~Myr. 
\\However, NGC~5044 is an exception. The difference in time between the two most recent outbursts in NGC~5044 is of $\leq$1 Myr, meaning that the thin disk phase in this system should have lasted only a few $10^{4}-10^{5}$ yr (see also \citealt{2021ApJ...906...16S}). Thus, it is possible that other scenarios apply to this peculiar system. For example, \citet{2013Sci...339...49M} showed that while the innermost region of a thick accretion flow remains aligned with the BH spin axis over time, the outer accretion flow can be tilted by a large angle. This outer thick disk can gradually push away and deflect the jet, which can eventually bend by up to 90$^{\circ}$ and continue to propagate orthogonal to the SBMH spin axis. Such deflection occurs at 100 gravitational radii ($r_{g}$), and by $300 \,r_{g}$ the orthogonal deflection is completed. 
For NGC~5044 ($M=2\times10^{8}$~M$_{\odot}$ and $1'' = 150\,\text{pc}$, see \citealt{2021ApJ...906...16S}), resolving such scales with radio observations would require a resolution of $\leq0.5$ mas. If any outer thick disk is deflecting the jets in NGC~5044, this can only be confirmed with radio observations that have a resolution of an order of magnitude higher than those currently available. 
\\Overall, we are unable to pinpoint the mechanism that can exhaustively explain the observed properties of the jet reorientation mechanism: the re-alignment of the jets can be fast (down to $\leq1$~Myr) and nearly orthogonal changes in position angle seem favored. If any temporary highly-efficient accretion phase has driven jet reorientation in the past, it must have been rather short lived ($\leq$ a few Myr). The solution may be provided in the future by models, simulations and observations that can account for the properties of AGN outbursts on scales that span several orders of magnitude (between 10$^{-4}$ pc to 10$^{4}$ pc).

\section{Conclusions} \label{sec:concl}
In this work we presented the analysis of VLBA and $Chandra$ data of 16 cool core galaxy clusters and galaxy groups, to understand how common jet and cavities misalignments are and which causes may be responsible for such mismatch. Below we report our main findings. 
\begin{itemize}[leftmargin=*,noitemsep]
    \item For each system we measured the jet position angle  ($\Phi_{\perp\text{LOS}}$) and the cavity position angle ($\Theta_{\text{cav}}$) in the plane of the sky. Using the full sample and selected subsets, we found that between 30\% to 38\% of the jet -- cavity pairs are misaligned by more than $\Delta\Psi~=~\Phi_{\perp\text{LOS}}~-~\Theta_{\text{cav}}~=~45^{\circ}$. 
    \item By measuring the jet inclination angle along the line of sight ($\Phi_{\parallel\text{LOS}}$) we determined that projection effects may bias our measurement of $\Delta\Psi$ in around half of the sample. Restricting the analysis to the AGN whose jets lie in the plane of the sky, we still found a fraction of $\sim30\%$ of misaligned systems, supporting the above conclusions.
    \item We retrieved information on the dynamical state of the host clusters and groups, finding that all are examples of disturbed systems. Even though we cannot exclude that gas motions may have perturbed X-ray cavities away from their original trajectory in some systems, the ubiquity of disturbances excludes the possibility that a perturbed environment plays the major role in producing large misalignments. 
    \item The distribution of $\Delta\Psi$ appears to be the combination of a Gaussian centered at $0^{\circ}$ with $\sigma = 27.6^{\circ}\pm1.6^{\circ}$ plus a peak around $\Delta\Psi\sim75^{\circ}$. We conclude that, at 2.5$\sigma$ significance, large misalignments are favored over small ones. 
    \item By considering the age of the X-ray cavities, we find that larger misalignments ($\Delta\Psi>75^{\circ}$) do not occur on longer timescales, with the shortest time difference being around 1 Myr in NGC~5044. Considering our findings and reviewing previous literature results, we cautiously propose that in addition to the spin axis of the SMBH, the geometry of the accretion flow may influence the direction of jet propagation. 
\end{itemize}

%% IMPORTANT! The old "\acknowledgment" command has be depreciated. It was
%% not robust enough to handle our new dual anonymous review requirements and
%% thus been replaced with the acknowledgment environment. If you try to 
%% compile with \acknowledgment you will get an error print to the screen
%% and in the compiled pdf.
%% 
%% Also note that the akcnowlodgment environment does not support long amounts of text. If you have a lot of people and institutions to acknowledge, do not use this command. Instead, create a new 
\section{Acknowledgments}
%\begin{acknowledgments}
%\textit{Acknowledgments}
We thank the referee for their useful comments on our work. FU thanks the Smithsonian Astrophysical Observatory for the hospitality and support during his visit. FU acknowledges support provided by the Marco Polo program. The authors acknowledge support for this work provided by the National Aeronautics and Space Administration (NASA) through Chandra Award Number GO1-22125X issued by the Chandra X-ray Center, which is operated by the Smithsonian Astrophysical Observatory for and on behalf of the National Aeronautics Space Administration under contract NAS8-03060. Basic research at the Naval Research Laboratory (NRL) is supported by 6.1 Base funding. WF acknowledges support from the Smithsonian Institution, the Chandra High Resolution Camera Project through NASA contract NAS8-03060, and NASA Grants 80NSSC19K0116, GO1-22132X, and GO9-20109X. The National Radio Astronomy Observatory is a facility of the National Science Foundation operated under cooperative agreement by Associated Universities, Inc. This work made use of the Swinburne University of Technology software correlator, developed as part of the Australian Major National Research Facilities Programme and operated under licence. This research has made use of data obtained from the Chandra Data Archive and the Chandra Source Catalog, and software provided by the Chandra X-ray Center (CXC) in the application packages CIAO and Sherpa.
%\end{acknowledgments}

%% To help institutions obtain information on the effectiveness of their 
%% telescopes the AAS Journals has created a group of keywords for telescope 
%% facilities.
%
%% Following the acknowledgments section, use the following syntax and the
%% \facility{} or \facilities{} macros to list the keywords of facilities used 
%% in the research for the paper.  Each keyword is check against the master 
%% list during copy editing.  Individual instruments can be provided in 
%% parentheses, after the keyword, but they are not verified.

\vspace{5mm}
\facilities{NRAO,CXO}

%% Similar to \facility{}, there is the optional \software command to allow 
%% authors a place to specify which programs were used during the creation of 
%% the manuscript. Authors should list each code and include either a
%% citation or url to the code inside ()s when available.

\software{\texttt{astropy} \citep{2013A&A...558A..33A,2018AJ....156..123A}, \texttt{APLpy} \citep{2012ascl.soft08017R}, \texttt{Numpy} \citep{2020Natur.585..357H}, \texttt{CIAO} \citep{2006SPIE.6270E..1VF}, \texttt{AIPS} \citep{2003ASSL..285..109G}, \texttt{DIFMAP} \citep{1994BAAS...26..987S}, \texttt{SHERPA} \citep{2021zndo...5554957B}.
          }

%% Appendix material should be preceded with a single \appendix command.
%% There should be a \section command for each appendix. Mark appendix
%% subsections with the same markup you use in the main body of the paper.

%% Each Appendix (indicated with \section) will be lettered A, B, C, etc.
%% The equation counter will reset when it encounters the \appendix
%% command and will number appendix equations (A1), (A2), etc. The
%% Figure and Table counter will not reset.
%\clearpage
\appendix
\section{Alternative methods to measure jets position angles on VLBA scales}\label{app:altmethod}
\begin{figure*}[!ht]
   \centering
   \includegraphics[width=1.0\linewidth]{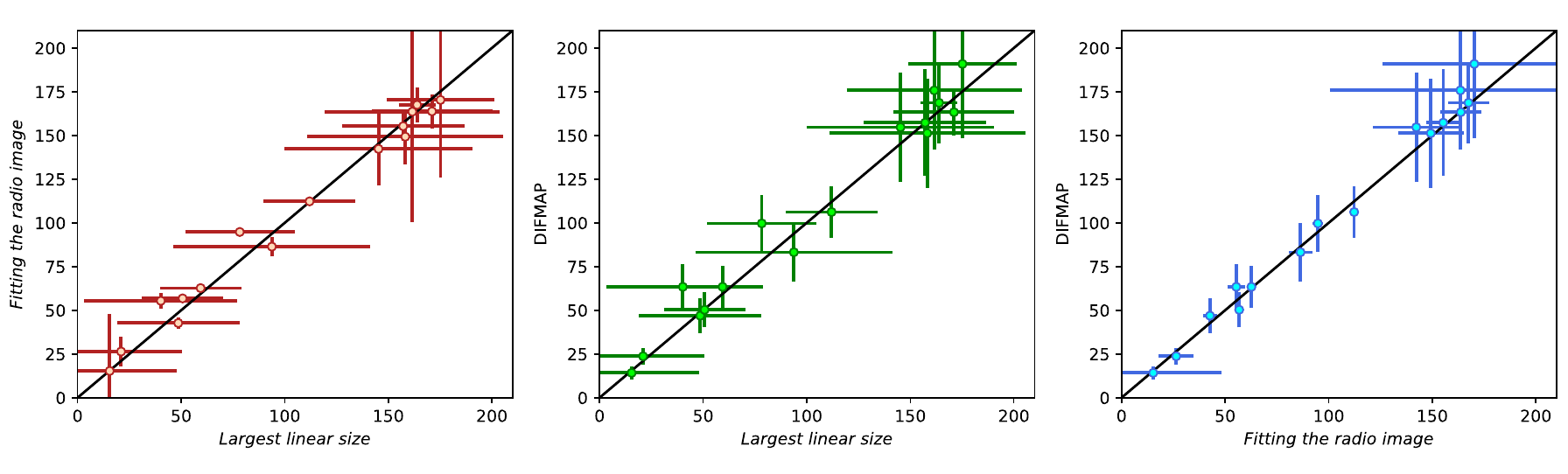}
      \caption{Comparison between the three methods we tested to measure the position angle of the radio jet in the plane of the sky ($\Psi_{\perp\text{LOS}}$). In the three panels, the black line represents the bisector of the plot and not a fit to the points. See Appendix \ref{app:altmethod} for details.
              }
         \label{fig:comparison-methods}
\end{figure*}

In this Appendix we show the two other methods we adopted to measure the position angle of the jets in a plane perpendicular to the line of sight.
\begin{itemize}
    \item \textit{Largest linear size}: we filtered the VLBA images to exclude all the pixels with a flux density lower than 5$\sigma$. Then, we computed the distance of each pixel from all the others. The two most distant pixels set the largest linear size of the radio source. The jet position angle is thus measured as the slope of the line that connects the two most distant pixels. The uncertainty on the position angle is the relative angular width of the beam with respect to the linear size, meaning that larger sources have smaller relative uncertainties. A potential weakness of this method is that in radio images the pixels are usually not independent, as the real resolution element is the beam, not the individual pixel. Local substructures (smaller than than the beam) may generate large differences in position angle for small differences in distance between adjacent pixels. 
    \item \textit{Fitting the radio image}: assuming that in the simplest configuration a core$+$jet morphology can be described as a straight line connecting two points (the core and the jet), we fitted the coordinates of the pixels (above 5$\sigma$) in the VLBA images with a straight line, thus treating the image as a scatter plot. Following this analogy, the jet position angle is the slope of the best-fit linear regression. During fitting, the pixels were weighted by their flux densities. Again, a potential weakness is that the pixels are not independent (there is the effect of the beam). Additionally, the assumption of a single line describing a core$+$jet morphology may not be accurate for highly resolved sources. 
\end{itemize}
In Fig. \ref{fig:comparison-methods} we show the comparison between the \textit{Largest linear size}, the \textit{Fitting the radio image}, and the \texttt{DIFMAP} methods.
Reassuringly, we find that the three methods provide consistent results. We note that the scatter of the points around the bisector of the plot is smaller than the errorbars: this is likely caused by the relatively large uncertainties that we assigned to the \textit{Largest linear size} and the \textit{Fitting the radio image} methods being conservatively overestimated. We decided to continue the analysis with the \texttt{DIFMAP} results because of its advantages: first, the beam shape is taken into account when fitting components to the visibilities; second, the \texttt{DIFMAP} method does not rely on cleaned images where pixel-scale artifacts may be present, as it works in the uv-domain.
%%%%%%%%%%%%%%%%%%%%%%%%FIGURE%%%%%%%%%%%%%%%%%%%%%%%
%%%%%%%%%%%%%%%%%%%%%%%%%%%%%%%%%%%%%%%%%%%%%%%%%%%%%
%%%%%%%%%%%%%%%%%%%%%%%%%%%%%%%%%%%%%%%%%%%%%%%%%%%%%
\section{Insights from VLBA data at other frequencies} \label{subsec:otherfreq}
\begin{figure*}[!ht]
   \centering
   \includegraphics[width=0.9\linewidth]{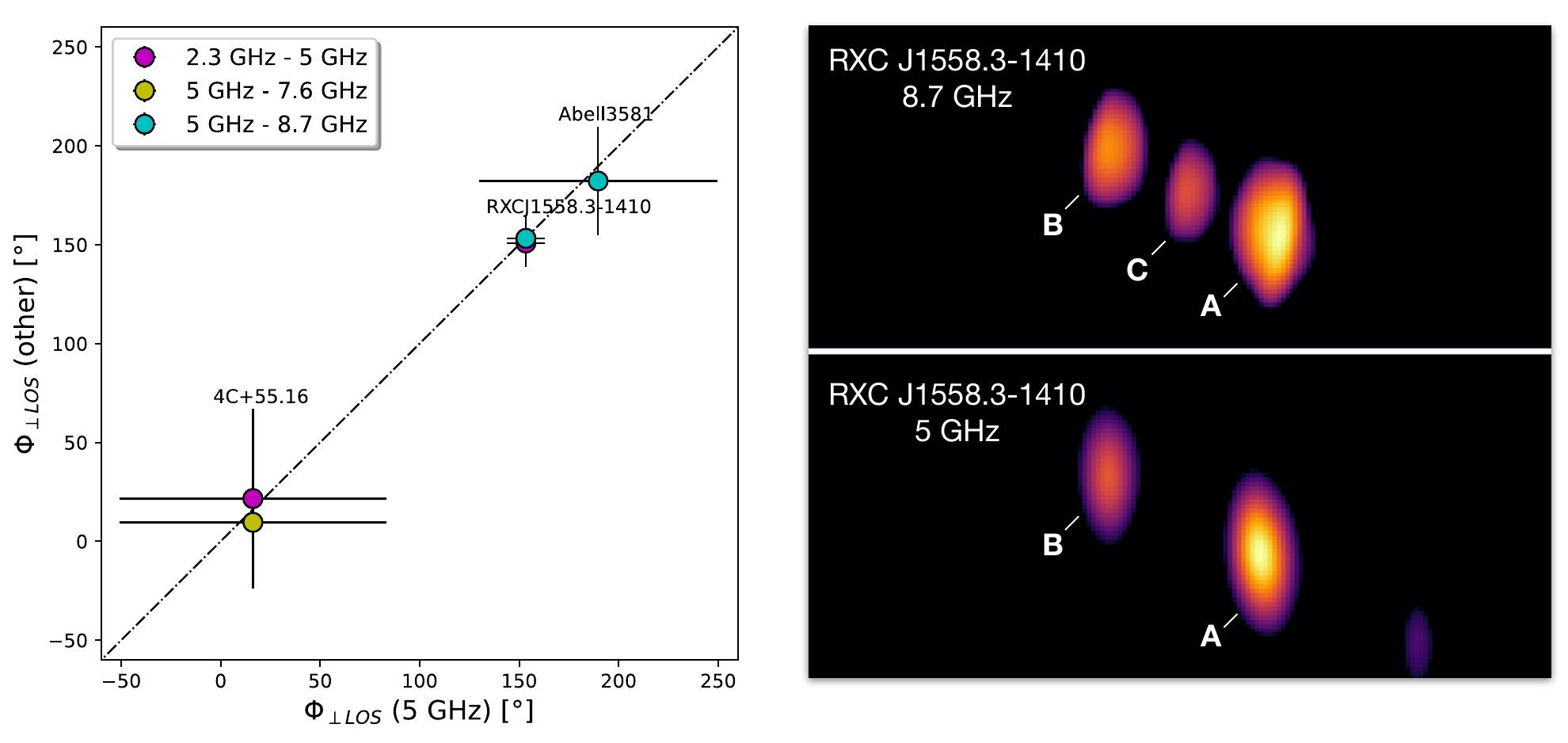}
      \caption{\textit{Left panel}: comparison between $\Phi_{\perp\text{LOS}}$ at 5~GHz and $\Phi_{\perp\text{LOS}}$ measured at other frequencies for the sources in our sample that are VLBA calibrators. The dashed line shows the $y=x$ bisector. \textit{Right panel}: VLBA images of RXC J1558.3-1410 at 8.7 GHz and at 5~GHz. The components identified in \texttt{DIFMAP} are labeled.
              }
         \label{fig:multifreq}
\end{figure*}

Three sources in our sample are VLBA calibrators\footnote{\url{https://obs.vlba.nrao.edu/cst/}, \citet{2020A&A...644A.159C}.}, meaning that there are archival snapshot observations at multiple frequencies other than 5~GHz. These are 4C+55.16 (2.3~GHz and 7.6~GHz), RXC~J1558.3-1410 (2.3~GHz and 8.7~GHz), and Abell~3581 (8.7 GHz). We used the archival observations of these three systems to validate our method of measuring $\Phi_{\perp\text{LOS}}$ from the 5~GHz maps. In practice, we applied the same fitting procedure in \texttt{DIFMAP} described in Sect. \ref{sec:measureangle}. The left panel of Fig. \ref{fig:multifreq} shows the results. It is clear that our measurements of $\Phi_{\perp\text{LOS}}$ are consistent between the different frequencies, thus strengthening our results. Radical changes in jet position angle with frequency are known (e.g., \citealt{2022ApJ...934..145I}), but these are typically found when not only the frequency but also the observed spatial scales change (e.g. from the parsec to the tens of parsecs scales). As the VLBA data at different frequencies probe relatively similar scales, it is likely that the observations used in this work do not suffer from this effect.
\\It is interesting to discuss the case of RXC~J1558.3-1410, which unveils a limitation of our work in terms of angle w.r.t. the line of sight, $\Phi_{\parallel\text{LOS}}$. For this AGN, the data at 8.7 GHz reveal a bright, unresolved component (labeled 'C') that is undetected at 5~GHz (see Fig. \ref{fig:multifreq}, right panel). If this component was a knot of the jet with a relatively steep spectral index ($\alpha\leq-0.5$), it should have been detectable at 5~GHz. Its non-detection suggests that it may be characterized by an inverted spectral index ($\alpha\geq0$), which is typical of the synchrotron self-absorbed cores of compact radio sources (e.g., \citealt{sadler2016}). While from the 5~GHz data alone we identified the core of the AGN as the brightest component, it is likely that the true core is located, in fact, between the two components detected at 5~GHz. While this is not an issue when measuring $\Phi_{\perp\text{LOS}}$, as the whole structure remains aligned, it biases our measurement of $\Phi_{\parallel\text{LOS}}$. Indeed, using the 5~GHz data alone we would mis-classify the jet and the counter-jet in this system. Component A in the 5~GHz map, previously classified as the core, has to be considered as the jet, while component B, previously classified as jet, is in fact the counter-jet. Incidentally, for RXC~1558.3-1410, the old and new $\Phi_{\parallel\text{LOS}}$ remain consistent with each other (changing from $39.8^{\circ}\pm11.8^{\circ}$ to $25.1^{\circ}\pm6.7^{\circ}$). However, this system highlights a limitation that may affect our measurements of $\Phi_{\parallel\text{LOS}}$, especially for one-sided sources. We verified the absence of an hidden core in all the other objects of our sample with available archival data at high frequency. For the remaining one-sided sources in our sample (ClG~J1532+3021, Abell~478, Abell~1664, NGC~6338, IC~1262, and NGC~5098) no data at other frequencies exist, but we plan to observe them with deeper and multi-frequency VLBA observations in the future.
\\Ultimately, we note that we are currently unable to measure the spectral index of the different components, which would benefit their classification as jets or core. The available multi-frequency data were not acquired close in time, meaning that flux variability at parsec scales could strongly bias the results of any spectral index study. Obtaining coeval multifrequency observations of our sample is a future perspective of this work.
%%%%%%%%%%%%%%%%%%%%%%%%FIGURE%%%%%%%%%%%%%%%%%%%%%%%
%%%%%%%%%%%%%%%%%%%%%%%%%%%%%%%%%%%%%%%%%%%%%%%%%%%%%
\section{RXC~J1558.3-1410: X-ray morphological analysis}\label{app:r1558}
\begin{figure*}[!ht]
   \centering
   \includegraphics[width=0.9\linewidth]{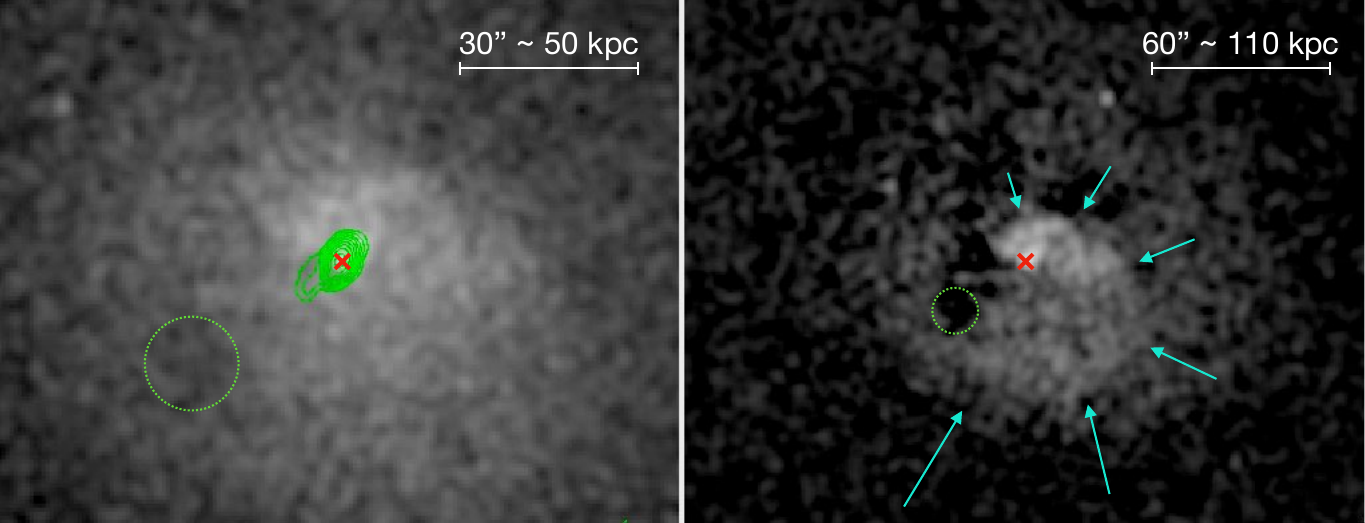}
      \caption{$Chandra$ images of RXC~J1558.3-1410. \textit{Left panel:} 0.5 -- 2 keV counts image, smoothed with a Gaussian of kernel size 3$''$. Green contours are the 1.4 GHz radio emission seen in archival VLA data, 
      %(project ID VLA:XH02022)
      starting from 3 $\times\sigma$ (with $\sigma=0.4$ mJy/beam, resolution of 2$''$) and increasing in steps of 2. \textit{Right panel:} residual $Chandra$ image, smoothed with a Gaussian of kernel size 5$''$. Cyan arrows show the spiral morphology of the ICM. In both panels, the green dashed circle shows the identified X-ray cavity, while the red cross shows the position of the central AGN.
              }
         \label{fig:r1558-appendix}
\end{figure*}

As reported in the main text, RXC~J1558.3-1410 lacks a dedicated X-ray study of the ICM morphology. To verify the existence of X-ray cavities in this system, we inspected both the $Chandra$ 0.5 -- 2 keV counts image and the residual image. In both of them, a depression in surface brightness is visible east of the center (see Fig. \ref{fig:r1558-appendix}). The depression is nearly circular, and it is located at $\sim20'' = 36$ kpc from the center. The signal-to-noise ratio (SNR) of this feature was measured in the $Chandra$ counts image by comparing the number of net counts within this region, $C_{c}$, with the average number of counts in similar regions with the same size at the same distance from the center, $C_{s}$, through the following expression: 
\begin{equation}
    \text{SNR}\,\, = \frac{|C_{c}-C_{s}|}{\sqrt{C_{c}+C_{s}}}
\end{equation}
For the depression shown in Fig. \ref{fig:r1558-appendix} (left panel) we measure a SNR $= 4.8$ (with $C_{c} = 892$ and $C_{s} = 1107$), which allows us to identify the feature as a reliable structure. Other depressions that are visible in the residual image do not represent statistically significant drops in the counts image (either they represent small deficits, around 5\%, or the SNR is lower than 3). Furthermore, we note that based on archival VLA data at 1.4 GHz (see \url{http://www.vla.nrao.edu/astro/nvas/}) there is evidence for radio emission extended towards this depression. This supports the classification of the X-ray depression as an AGN-inflated cavity in the ICM. 
\\ We also note that on larger scales, the surface brightness distribution has a spiral morphology, with a bright edge west and south-west of the center (see Fig. \ref{fig:r1558-appendix}, right panel). This is typical of clusters experiencing sloshing of the hot gas, which is thus likely present also in RXC~J1558.3-1410. The spiral structure is wrapped around the cluster core, and is visible up to a distance of $\sim$100 kpc from the center.
%%%%%%%%%%%%%%%%%%%%%%%%FIGURE%%%%%%%%%%%%%%%%%%%%%%%
%%%%%%%%%%%%%%%%%%%%%%%%%%%%%%%%%%%%%%%%%%%%%%%%%%%%%

\section{Radio data and \texttt{DIFMAP} results}\label{app:difmap}
In this Appendix we present the results of fitting the VLBA data for our sample in \texttt{DIFMAP}. Tab. \ref{tab:difmapcomp} reports, for each system, the properties of the radio image shown in Fig. \ref{fig:vlbaima} and the parameters of the components used to fit the visibilities in the uv-domain.
%\begin{center}
\setlength{\tabcolsep}{8pt}
\begin{table*}[ht]
        \centering
        \renewcommand{\arraystretch}{0.9}
		%\begin{longtable}{l|c|c|c|c|c|c|c|c|c}
		\caption{Properties of the VLBA images shown in Fig. \ref{fig:vlbaima} and results of fitting the visibilities in \texttt{DIFMAP}. (1) Name of the system; (2) size and position angle (from the north axis) of the restoring beam; (3) noise of the VLBA image; (4 -- 10) \texttt{DIFMAP} results: label of the component, flux density, distance from phase center (polar coordinates), angle from the phase center (polar coordinates, with $\theta$ starting from the north axis), major semi-axis, axes ratio, orientation of the component (from the north axis). When $a = 0$ the component is a $\delta$-function. When $b/a = 1$ the component is a circle. When $a\neq0$ and $b/a\neq1$ the component is an ellipse with orientation $\psi$.}
	    \label{tab:difmapcomp} 
        \begin{tabular}{l|c|c|c|c|c|c|c|c|c}
			\hline
			Cluster &  Beam, PA & $\sigma$ & Comp. & Flux & r & $\theta$ & a & b/a & $\psi$ \\
			  & [mas$\times$mas, $^{\circ}$] & [mJy/beam] & & [mJy] & [mas] & [deg] & [mas] & & [deg]\\
			\hline
            Abell~2390   & $3.5\times1.2$, 0.6 &  0.1  & A &  118 &0.002 &-160.2 &  0 &  1.0 &  -8.3 \\
                      &                     &       & B &  7.6 & 4.68 & 178.6 &  1.2 &  1.0 &     0 \\
                      &                     &       & C & 57.5 & 3.43 &  -7.0 &    0 &  1.0 &     0 \\
                      &                     &       & D & 22.8 & 22.2 & -24.1 & 14.9 &  1.0 &     0 \\                      
		
            \hline
	        ClG~J1532+3021  & $5.0\times1.4$, 25.1& 0.05  & A &  6.2 & 11.40& 108.5 &  0 &  1.0 &     0 \\
                      &                     &       & B &  1.5 &  8.59& 117.8 &  1.8 &  1.0 &     0 \\
		
            \hline
			4C+55.16  & $3.5\times1.7$, 4.4 &  5.0  & A & 2136 & 0.67 &  83.1 &  0 & 1.0 & 0 \\
                      &                     &       & B & 1561 & 1.58 & -64.3 &  8.4 & 0.37 & -51.6 \\

            \hline
            Abell~478     & $4.0\times1.6$, 11.4& 0.05  & A &  5.4 & 11.21& 106.4 &  0.8 &  1.0 &     0 \\
                      &                     &       & B &  0.9 & 14.28& 117.6 &    0 &  1.0 &     0 \\
		
            \hline
            RX~J1447.4+0827  & $3.2\times1.1$, -6.0& 0.05  & A &  8.1 & 86.69&  84.7 &  0.4 &  1.0 &     0 \\
                      &                     &       & B &  5.6 & 86.11&  83.2 &  0.6 &  1.0 &     0 \\
                      &                     &       & C &  3.8 & 87.39&  86.0 &  1.0 &  1.0 &     0 \\                      
		
            \hline
            Abell~1664    & $3.3\times1.0$, -3.6& 0.05  & A &  9.5 &500.45& 144.5 &  0 &  1.0 & 0 \\
                      &                     &       & B &  2.5 &503.46& 144.9 &  1.3 &  1.0 &     0 \\				
		
            \hline
			RXC~J1558.3-1410 
                     
			          & $3.5\times1.2$, 0.5 &  1.5  & A &  287 & 0.04 & -24.0 &  1.2 &  1.0 &     0 \\
                      &                     &       & B &  117 & 7.74 &  63.9 &  1.2 &  1.0 &     0 \\
                      %&                     &       %& C & 39.6 & 7.72 &-116.7 &    0 &  1.0 &     0 \\                       

            \hline
            ZwCl~8276 & $3.5\times1.2$, -9.0& 0.08  & A & 72.8 &  0.11& -11.4 &    0 &  1.0 &     0 \\
                      &                     &       & B &  5.6 &  1.67&  -3.2 &    0 &  1.0 &     0 \\
                      &                     &       & C & 15.7 &  1.09& 174.8 &    0 &  1.0 &     0 \\	
		
            \hline
            Abell~496     & $3.6\times1.2$, 1.5 & 0.06  & A & 44.3 & 69.49&  29.3 &  0.7 &  1.0 &     0 \\
                      &                     &       & B &  7.3 & 74.79&  30.6 &    0 &  1.0 &     0 \\
                      &                     &       & C &  5.3 & 62.84&  28.5 &    0 &  1.0 &     0 \\
		
            \hline
            ZwCl~235  & $3.3\times1.2$, -1.9& 0.07  & A & 16.0 & 65.54&-145.5 &  0.7 &  1.0 &     0 \\
                      &                     &       & B &  3.6 & 64.36&-146.0 &  1.2 &  1.0 &     0 \\
                      &                     &       & C &  5.3 & 66.63&-145.0 &  1.2 &  1.0 &     0 \\	
		
            \hline
            Abell~2052    & $3.3\times1.2$, 1.9 &  1.7  & A &  267 & 0.03 &  72.6 &  0.6 &  1.0 &  0 \\
                      &                     &       & B & 64.5 & 4.56 & 170.8 &  1.9 &  1.0 &     0 \\
                      &                     &       & C & 57.5 & 4.06 & -17.2 &  1.9 &  1.0 &     0 \\                      
            \hline
			Abell~3581    & $4.1\times1.2$, 0.6 &  0.3  & A & 99.2 & 0.04 & -22.5 &  0.8 &  1.0 & 0 \\
                      &                     &       & B &  128 & 0.34 &  58.5 &  4.8 &  1.0 &     0 \\
                      &                     &       & C & 35.6 & 7.25 &. 97.9 &  3.0 &  1.0 &     0 \\     
                    
            \hline
            NGC~6338  &$4.3\times1.3$, -10.6& 0.1  & A & 19.9 &  0.09&   1.8 &  0.8 &  1.0 &     0 \\
                      &                     &       & B &  5.3 &  4.12&-155.8 &  3.1 &  1.0 &     0 \\

            \hline
			IC~1262$^{\ast}$   &$5.3\times3.8$, -11.9& 0.04  & A &  3.1 &395.39&117.8 &  0.7 &  1.0 &  0 \\
                               &                     &       & B &  0.3 &397.27&118.9 &  4.2 &  1.0 &      0 \\	
		
			\hline
			NGC~5098  & $5.0\times1.6$, 15.5& 0.05  & A &  4.7 & 34.19&  35.6 &  1.7 &  1.0 &     0 \\
                      &                     &       & B &  1.1 & 33.12&  39.2 &  0.7 &  1.0 &     0 \\				
		
            \hline
			NGC~5044  & $3.9\times1.3$, -4.4&  0.03 & A & 16.5 &147.03& -67.8 &  1.9 &  1.0 &     0 \\
                      &                     &       & B &  4.8 &150.99& -70.5 &  6.6 &  1.0 &     0 \\
                      &                     &       & C &  0.8 &149.64& -64.5 &    0 &  1.0 &     0 \\

            \hline
			
            \end{tabular}
            \tablecomments{$^{\ast}$ The radio map has been created by applying an uv-taper of 30~M$\lambda$ to the visibilities.}
		%\end{longtable}
        \end{table*}
		
%\end{center}

\section{Comments on individual sources}\label{app:comment}
\noindent \textbf{Abell~2390}: galaxy cluster at z = 0.23. The VLBA image shows a two-sided jet structure in the north-south direction, predominantly in the plane of the sky (consistent with \citealt{2006MNRAS.367..366A}). From the X-ray point of view, no unique view on the presence of cavities in this system is present in the literature. From a heavily smoothed and filtered image, \citet{2015ApSS.359...61S} identified a total of 4 X-ray cavities surrounding the central AGN. From our re-analysis of the X-ray data (using all the three available ObsIDs) we find compelling evidence for the existence of one of the X-ray cavities identified by \citet{2015ApSS.359...61S}, west of the radio core. Additionally, there are large X-ray cavities east -- west of the core, at a distance from the center of $\sim$200 kpc (see \citealt{2011xru..conf..241L}). These structures are filled by radio emission in the form of large lobes in LOFAR images (see \citealt{2020MNRAS.496.2613B}). Between the two generations of X-ray cavities and the inner jets we find a misalignment of almost 90$^{\circ}$, with projection effects playing at most a minor role. \\
\textbf{ClG~J1532+3021}: galaxy cluster at z = 0.362 studied by \citet{2013ApJ...777..163H}, who revealed the presence of sloshing motions and X-ray cavities. Two X-ray cavities are clearly visible east -- west of the radio core in the {\it Chandra} image. The VLBA observations reveal a relatively faint source with a small one-sided jet west of the core, nearly aligned with the X-ray cavities. \\
\textbullet \,\,\textbf{4C+55.16}: galaxy cluster at z=0.242. \citet{2011MNRAS.415.3520H} discovered two X-ray cavities on opposite sides of the radio core. The authors noted a surface brightness edge, interpreted as a cold front, south of the cluster's center.  \citet{1985MNRAS.214...55W} revealed a pair of radio lobes in the southeast -- northwest direction, that coincide with the X-ray cavities. On VLBA scales we find resolved extended emission that is well fitted in \texttt{DIFMAP} by two components, that we identify as the core and a jet. Such interpretation is consistent with the EVN images at 5~GHz published by \citet{1985MNRAS.214...55W}. The northwest cavity is aligned with the radio jet, while the southeast one is misaligned. It is important to note that projection effects may be relatively important in this system, where $\Psi_{\parallel\text{LOS}}\sim 60^{\circ}$. \\
\textbf{Abell~478}: galaxy cluster at z = 0.088. \citet{2003ApJ...587..619S} and \citet{2014ApJ...781....9G} reported the presence of small X-ray cavities in the north -- south direction, and the disturbed state of the cluster. From the VLBA data we recover a one-sided source, misaligned by $\sim$35$^{\circ}$ from the X-ray cavities. The source has a jet-counterjet ratio of roughly 6. Thus, projection effects may be important in this system. Overall, given the mild difference in P.A. between jet and X-ray cavities, and projection effects, it is unclear whether this system can be classified as aligned or misaligned. \\
\textbf{RX~J1447.4+0827}: galaxy cluster at z = 0.376. The presence of small X-ray cavities and sloshing signatures in this object has been reported by \citet{2020AJ....160..103P}. On VLBA scales, the source is clearly resolved in a two-sided structure oriented northwest -- southeast. The jets are aligned with the cavities on larger scales, and the low jet -- counterjet ratio (1.1) indicates that projection effects are likely negligible. \\
\textbf{Abell~1664}: galaxy cluster at z = 0.128. \citet{2019ApJ...875...65C}, revealed the presence of two pairs of X-ray cavities (the inner pair in the east -- west direction, the outer pair in the southeast -- northwest direction), and of a disturbed ICM. The VLBA data are difficult to interpret: there is extended, diffuse emission south of the brightest component, however its morphology is not straightforwardly reminiscent of a jet. By interpreting the radio source as a core + jet system, the position angle of the AGN jet is misaligned w.r.t. the cavities. The source is also apparently one-sided, with an angle w.r.t. the line of sight of $\sim50^{\circ}$.  \\
\textbf{RXC~J1558.3-1410}: galaxy cluster at z=0.097. As no dedicated studies of the {\it Chandra} observations of this object exist in the literature, we analyzed the data (see Appendix \ref{app:r1558}). We identify a cavity southeast of the AGN (the depression is significant at a signal-to-noise ratio of SNR=6). We also identify a positive residual spiral from the residual {\it Chandra} image, that we attribute to sloshing of the ICM. Archival VLA observations at 1.4 GHz reveal a jet feature headed towards the depression, supporting its cavity classification. Being a VLBA calibrator, this radio galaxy has multi-frequency coverage in the archive. An inspection of the 8.7 GHz observations reveals a two-sided source. Based on the 5~GHz observations only, the brightest component (A) may be misinterpreted as the radio core. However, the 8.7 GHz map reveals a third component left of the brightest one. This third component is likely the true core, self-absorbed (thus invisible) at lower frequencies (see Appendix \ref{subsec:otherfreq} for details). \\
\textbf{ZwCl~8276}: galaxy cluster at z = 0.076. Several structures in the ICM support the idea that the cluster is experiencing sloshing (see \citet{2013A&A...555A..93E}); X-ray cavities (in the north -- south direction) have been identified by \citet{2013A&A...555A..93E}. From our $Chandra$ images, and guided by the recent LOFAR observations of this cluster \citep{2020MNRAS.496.2613B}, we identify a third, outer and larger X-ray cavity northeast of the AGN. In the VLBA images there are compelling evidence for barely resolved north -- south extensions, aligned with the inner X-ray cavities.  \\
\textbf{Abell 496}: galaxy cluster at z = 0.033. Several studies have focused on the dynamical status of the ICM in this object, given the clear sloshing spiral visible in X-ray observations (e.g., \citet{2012MNRAS.420.3632R,2014A&A...570A.117G}). However, no dedicated studies on AGN feedback in this system are available, despite the clear presence of multiple, aligned X-ray cavities in the northeast -- southwest direction, coincident with extended radio emission from the central AGN (see \citealt{giacintucci2016}; Giacintucci et al. in prep.). The VLBA data reveal a two-sided structure in the same direction, suggesting that over 3 cycles of radio activity the AGN's jet axis has remained nearly constant. From the jet-counterjet ratio we exclude that projection effects play a major role. \\
\textbf{ZwCl~235}: galaxy cluster at z = 0.083. \citet{ubertosi_23AA} revealed two radio-filled cavities in the core of the cluster oriented in the northeast -- southwest direction, and of a sloshing spiral wrapped around the core. The VLBA data unveil a two-sided radio source nearly aligned with the X-ray cavities, mainly in the plane of the sky. \\
\textbf{Abell 2052}: well-studied galaxy cluster at z = 0.035. The ICM in this object shows multiple X-ray cavities and shock fronts around the central AGN, signals of multiple outbursts of the radio galaxy (see \citet{2011ApJ...737...99B} and references therein). The two pairs of X-ray cavities are roughly oriented in the north-south direction. The VLBA reveal a two-sided source extended in the north-south direction. The position angle of the jets is consistent with the results obtained from multifrequency imaging by \citet{2004A&A...422..515V}. Overall, the jets and the cavities are roughly aligned in this system, where projection effects are likely negligible. \\
\textbf{Abell 3581}: galaxy group at z=0.022. \citet{2005MNRAS.356..237J,2013MNRAS.435.1108C} reported the presence of a sloshing ICM spiral wrapped around the core and two clear cavities east -- west of the central AGN, filled by the radio lobes of the radio galaxy. A third X-ray cavity is found beyond the west cavity and roughly aligned with the inner pair. From the VLBA data (5~GHz and 8.7 GHz) we observe a strongly one-sided source, with a core-jet morphology in the east-west direction, in good agreement with the position of the X-ray cavities. \\
\textbf{NGC~6338}: merging galaxy group at z=0.027. The multifrequency study of this object has been carried out by \citet{2019MNRAS.488.2925O}, who found radio-filled X-ray cavities in both groups. For our analysis we focus on the southern, larger subgroup, that has two inner X-ray cavities oriented northeast -- southwest, and at least one X-ray cavity on larger scales, west of the core (see also \citealt{2020MNRAS.496.2613B,2023ApJ...948..101S}). The VLBA data reveal a core $+$ jet morphology. The source appears one-sided, although at a level of 3$\sigma$ ($\sim$0.4 mJy/beam) there are indications for a counterjet. The radio source is oriented northeast -- southwest, thus being aligned with the inner X-ray cavities but misaligned w.r.t. the older, outer outburst. \\
\textbf{IC~1262}: galaxy group at z = 0.033. The hot gas in this object shows several substructures (ripples, edges and depressions). There are two inner X-ray cavities (oriented north --south) and an outer, larger X-ray cavity south of the center (see \citealt{2019ApJ...870...62P}). The VLBA data reveal an extended feature south of the core, that is recovered only by tapering the visibilities during imaging. The position angle of the jet is in relatively good agreement with that of the inner and outer X-ray cavities.\\
\textbf{NGC~5098}: galaxy group at z=0.036. \citet{2009ApJ...700.1404R} reported the presence of two X-ray cavities filled by radio emission and mainly oriented in the north -- south direction. The group also shows signs of minor merger with a galaxy, that is causing sloshing of the IGrM. On VLBA scales we find a core dominated source with a small jet southeast of the core, nearly aligned with the X-ray cavities. \\
\textbf{NGC~5044}: well-studied galaxy group at z=0.009. Several works reported the presence of multiple X-ray cavities in the intragroup medium of this object, with the different pairs approximately aligned along the southeast -- northwest direction (see \citealt{2021ApJ...906...16S} and references therein). The VLBA data reveal a two-sided core-jet structure rotated by $\sim$90$^{\circ}$ with respect to the successive cavity pairs (see also \citealt{2021ApJ...906...16S}). Based on the jet-counterjet ratio the inner jets, the difference in position angle of the jets and the cavities is real and not strongly influenced by projection effects.

%% For this sample we use BibTeX plus aasjournals.bst to generate the
%% the bibliography. The sample631.bib file was populated from ADS. To
%% get the citations to show in the compiled file do the following:
%%
%% pdflatex sample631.tex
%% bibtext sample631
%% pdflatex sample631.tex
%% pdflatex sample631.tex

\bibliography{sample631.bib}{}
\bibliographystyle{aasjournal}

%% This command is needed to show the entire author+affiliation list when
%% the collaboration and author truncation commands are used.  It has to
%% go at the end of the manuscript.
%\allauthors

%% Include this line if you are using the \added, \replaced, \deleted
%% commands to see a summary list of all changes at the end of the article.
%\listofchanges

\end{document}